\documentclass[letterpaper,twocolumn,10pt]{article}
\usepackage{usenix}
\usepackage{tikz}
\usepackage{amsmath}

\usepackage{filecontents}

\usepackage[english]{babel}
\usepackage{blindtext}

\usepackage{graphicx, xspace, enumitem, balance}

\usepackage[ruled,vlined]{algorithm2e}
\usepackage{algorithmic}
\usepackage{subfigure}

\newcommand{\paraspace}{\vspace{0.01in}}
\newcommand{\parab}[1]{\paraspace\noindent{\bf #1.}}

\newcommand{\fixme}[1]{\textbf{\textcolor{red}{#1}}}

\newcommand{\sys}{\textsf{FC-BGP}\xspace}

\newcommand{\fcshort}{\textsf{FC}\xspace}
\newcommand{\fcs}{\textsf{FCs}\xspace}

\newcommand{\first}{\textsf{(i)}\xspace}
\newcommand{\second}{\textsf{(ii)}\xspace}
\newcommand{\third}{\textsf{(iii)}\xspace}
\newcommand{\fourth}{\textsf{(iv)}\xspace}

\newcommand{\eg}{\emph{e.g.,}\xspace}
\newcommand{\ie}{\emph{i.e.,}\xspace}

\begin{document}

\date{}

\title{\Large \bf Secure Inter-domain Routing and Forwarding via \\ Verifiable Forwarding Commitments}


\author{
{\rm Xiaoliang \ Wang}\\
Capital Normal University \\
wangxiaoliang@cnu.edu.cn
\and
{\rm Zhuotao \ Liu}\\
Tsinghua University \\
zhuotaoliu@tsinghua.edu.cn
\and
{\rm Qi \ Li}\\
Tsinghua University \\
qli01@tsinghua.edu.cn
\and
{\rm Yangfei \ Guo}\\
Zhongguancun Laboratory \\
guoyangfei@zgclab.edu.cn
\and
{\rm Sitong \ Ling}\\
Tsinghua University \\
lingst21@mails.tsinghua.edu.cn
\and
{\rm Jiangou \ Zhan}\\
Tsinghua University \\
zjo23@mails.tsinghua.edu.cn
\and
{\rm Yi \ Xu}\\
Tsinghua University \\
yi-xu18@mails.tsinghua.edu.cn
\and
{\rm Ke \ Xu}\\
Tsinghua University \\
xuke@tsinghua.edu.cn
\and
{\rm Jianping \ Wu}\\
Tsinghua University \\
jianping@cernet.edu.cn
} 

\maketitle

\begin{abstract}
The inter-domain routing system is the most fundamental building block of Internet interconnection. However, it is vulnerable to various attacks. On the control plane, the de facto Border Gateway Protocol (BGP) does not have built-in mechanisms to authenticate routing announcements, so an adversary can announce virtually arbitrary paths to hijack network traffic; on the data plane, it is difficult to ensure that actual forwarding path complies with the control plane decisions. 
The community has proposed significant research to secure the routing system. Yet, existing secure BGP protocols (\eg BGPsec) make a positive contribution to improving the BGP security while facing the challenge of not being incrementally deployable. Meanwhile, existing path authorization protocols are not compatible with the current Internet routing infrastructure. 

In this paper, we propose \sys, the first secure Internet inter-domain routing system that can simultaneously authenticate BGP announcements and validate data plane forwarding in an efficient and incrementally-deployable manner. \sys is built upon a novel primitive, named Forwarding Commitment, to  certify an AS's \emph{routing intent} on its directly connected hops. 
We provide a rigorous security analysis and demonstrate that \sys achieves the same security guarantees as BGPsec while offering significantly more security benefits in case of partial deployment. Further, we implement a prototype of \sys and extensively evaluate it over a large-scale overlay network with 100 virtual machines deployed globally. The results demonstrate that \sys saves roughly 36\% of the overhead required to validate all BGP announcements, and meanwhile \sys introduces minimal overhead for building a globally consistent view on the authorized forwarding paths on data plane. 


\end{abstract}

\section{Introduction}

The fundamental cause of the path manipulation attacks in Internet inter-domain routing is that the de facto Border Gateway Protocol (BGP) does not have built-in mechanisms to authenticate routing announcements. As a result, an adversary can announce virtually arbitrary paths to a prefix while the network cannot effectively verify the authenticity of the route  announcements.  
In addition to the lack of control plane authentication, ensuring that the actual forwarding paths in the dataplane comply with the control plane decisions is also missing in today's inter-domain routing system. This fundamentally limits ASes from filtering unauthorized traffic~\cite{kim2014lightweight,legner2020epic}.  

The Internet community has conducted significant research in addressing above problems. Regarding the control plane, the community proposes to embed a form of authentication code (such as cryptographic signatures) inside the route announcement to make it self-verifiable. The most representative solutions given by Internet Engineering Task Force (IETF) are: origin authentication with the Resource Public Key Infrastructure (RPKI)~\cite{RPKI} and path validation through replacing BGP with BGPsec~\cite{BGPsec}.
Regarding the dataplane, the research community proposed path-aware Internet routing architecture~\cite{2010SCION,2009Pathlet}, to enforce various path authorization protocols, such as ICING~\cite{naous2011verifying}, OPT~\cite{kim2014lightweight}, and EPIC~\cite{legner2020epic}.
IETF also advocates source address validation architecture (SAVA)~\cite{wu2008source} and best practices to block packets with spoofed source addresses~\cite{DBLP:conf/icnp/WuRL07,wu2008source,ferguson2000defeating,baker2004ingress}. 

Yet, we recognize several key limitations in prior art. First, BGPsec is not incrementally deployable. It tightly couples the path authentication with the BGP path construction itself, where an AS is required to iteratively verify the signatures of each prior hop before extending the authentication chain with its own approval. As a result, a single legacy or malicious AS can terminate the authentication chain, preventing the downstream ASes from reinstating the authentication process. This imposes a legitimate deterrent to the ASes for adopting BGPsec because \emph{whether they can attest routing preference even for their directly connected hops is at stake}.
In addition, the performance hit introduced by BGPsec is significant because it has to authenticate the entire path even if only part of the hops is changed. 
Meanwhile, although path authorization~\cite{kim2014lightweight,legner2020epic} is native to path-aware Internet architecture, none of these protocols are fully compatible with BGP. This implies that unless the current Internet routing system would experience a fundamental paradigm shift towards path-aware routing, it is challenging to enforce path authorization in the inter-domain routing.

To address above challenges, we present \sys, an incrementally deployable security augment to the Internet inter-domain routing and forwarding. 
The key primitive in \sys is the Forwarding Commitment (\fcshort), which is a publicly verifiable code that certifies an AS's \emph{routing intent} on one of its directly connected hops, \ie an \fcshort  indicates whether the AS is willing to carry traffic for a specific prefix over the hop to a specific previous AS. 
Upon receiving a BGP announcement, if an AS decides to accept this route and extends the path to its (selected) neighbors, the AS commits its routing intent by generating a cryptographically-signed \fcshort.  
Therefore, downstream on-path ASes can validate the correctness of a BGP update by checking the \fcs associated with the individual hops on the AS-path. 
Because the \fcs are designed to be hop-specific and path-agnostic, a deployed AS can immediately certify its routing intent regardless of the deployment status of other ASes. This is fundamentally different from any path-level BGP authentication protocol (\eg BGPsec) where an on-path AS cannot approve any form of routing intent unless all on-path ASes are upgraded. 

Meanwhile, the flexibility of \fcs further enables efficient forwarding validation on the dataplane. Specifically, because the \fcs are self-proving, an AS can conceptually construct a certified AS-path using a list of consecutive per-hop \fcs, and then \emph{binds} its network traffic (identified by $[\textsf{src-AS}, \textsf{dst-AS}, \textsf{prefix}]$) to the path. 
This binding information essentially defines the authoritative forwarding path for the traffic $[\textsf{src-AS}, \textsf{dst-AS}, \textsf{prefix}]$. Therefore, by advertising the binding information globally, both on-path and off-path ASes are aware of the desired forwarding paths so that they can collaboratively discard the unwanted traffic that takes an unauthorized path. 
\sys designs a lightweight synchronization protocol to ensure that the global binding information is consistent.  

The contributions of this paper is as follows. 
\begin{itemize}[leftmargin=*]
\setlength\itemsep{0em}
    \item To the best of our knowledge, \sys is the first secure inter-domain routing system that can simultaneously authenticate BGP routing updates and validate dataplane forwarding in an efficient and incrementally-deployable manner. Crucially, \sys is built upon a unified primitive, named Forwarding Commitment (\fcshort), to enhance the security of control plane routing and dataplane forwarding. 
    
    \item Through rigorous security analysis, we demonstrate that \sys, although built upon hop-level \fcs, achieve the same security guarantees as BGPsec (a path-level authentication approach), while offering significantly more security benefits in case of partial deployment. 
    
    \item We further implement a prototype of \sys and extensively evaluate it in a large-scale overlay network with 100 virtual machines deployed across multiple continents. The results show that \first In a dynamic network such as the Internet, \sys reduces the overhead of validating all BGP announcements by approximately 36\%; \second \sys introduces minimal overhead for creating a globally consistent view of the authorized forwarding paths on the data plane, compared to simply relying on distributed consensus protocols such as PBFT. 
\end{itemize}

\section{Overview}


\subsection{Threat Model and Assumptions}
\label{subsec:Threatmodel}

We assume that ASes participating in \sys have access to an Internet-scale trust base, namely Resource Public Key Infrastructure (RPKI), that stores authoritative information about the mapping between AS numbers and their owned IP prefixes, as well as ASes' public keys. Also, we assume that there is no multi-path forwarding that is unknown to the source AS.

Given the above assumptions, we consider the following adversary. \first The adversary can intercept all the BGP update messages (also referred to as BGP announcements) in the network. \second On the control plane, the adversary can launch \emph{path manipulation attacks}. This means that the adversary will try to hijack traffic from the victim ASes by sending bogus BGP updates with a shorter path. \third On the data plane, the adversary can spoof source addresses and manipulate the forwarding path. \fourth The adversary will not engage in collusive attack, \ie we assume that two compromised ASes do not collude with each other. 

\subsection{Security Goals}
\label{subsec:design_goals}

\begin{figure*}
\centering
 \includegraphics[width=0.95\textwidth]{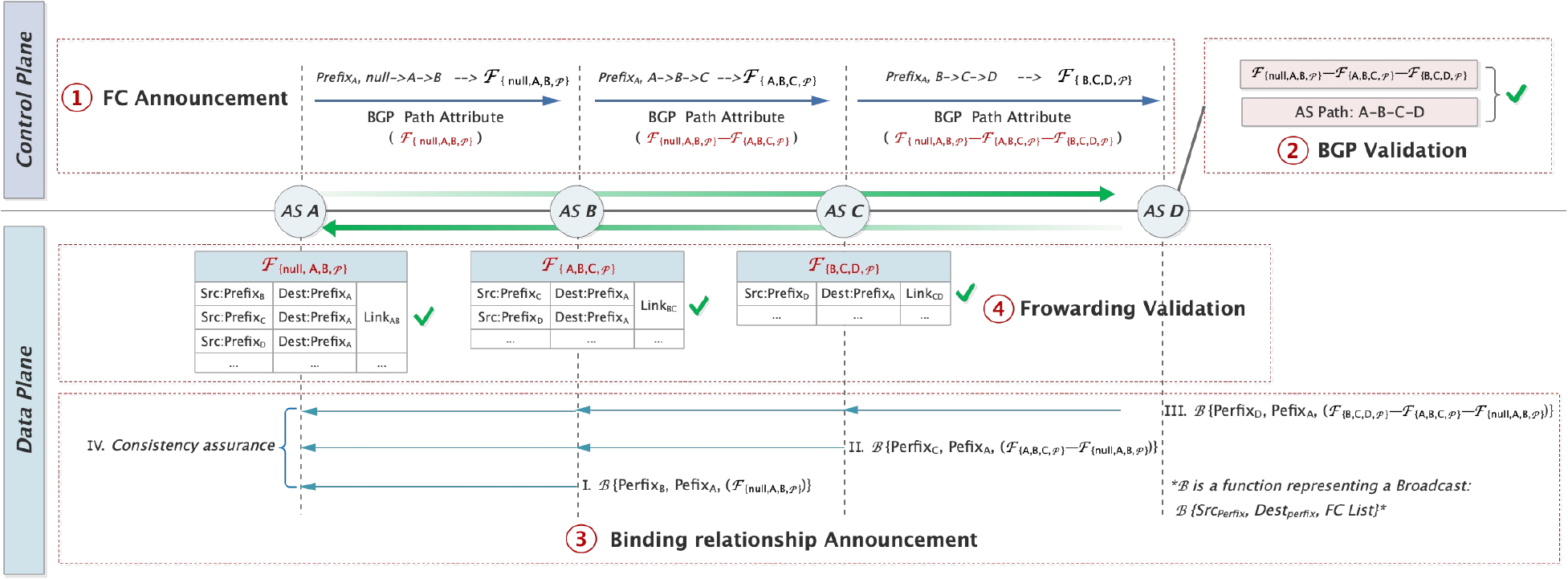}
 \caption{Overview of \sys.}
\label{example}
\end{figure*}

\sys is designed to enhance the security of both inter-domain routing and forwarding, even in case of partial deployment. 

\parab{Control Plane} 
In \sys, a BGP announcement is considered to be authenticated if the AS path specified in the announcement can be verified by the set of \fcs carried in the announcement. This means that we can find a properly signed \fcshort for every hop on the AS path. 
If all of the ASes in the network are upgraded to support \sys, then \sys can guarantee that any authenticated BGP announcement is \emph{actually advertised by the ASes on the AS-path}. In other words, it is infeasible for an adversary to claim that a non-exist AS-path is authenticated. 

In the case of partial deployment (\ie when only a certain percentage of ASes have upgraded to support \sys), we design \sys to provide more security benefits than BGPsec, which is currently being standardized by the IETF. 
At a high level, \sys enables greater partial deployment benefits because downstream upgraded ASes can use the sub-path (or pathlet) that have been authenticated by upstream upgraded ASes, even if the entire path is divided by latency ASes. 
In Section~\ref{sub:SecurityofCP}, we provide detailed analysis to demonstrate that: \sys can prevent the entire AS-path from being hijacked when the authenticated pathlet is sufficiently long (\ie when it does not require every AS on the path to be upgraded); and \second given the same deployment rate, \sys achieves a much lower path hijacking rate than BGPsec.

\parab{Data Plane} Denote an authenticated BGP path from AS $A$ to AS $N$ for prefix $\mathcal{P}$ as $\mathcal{P}_{A\mbox{-}N}$. 
Then the authorized forwarding path on the data plane for the traffic $T$ sent from AS $N$ to AS $A$ should be consistent with $\mathcal{P}_{A\mbox{-}N}$, \ie the traffic should flow from AS $N$ to AS $A$ in the opposite direction of $\mathcal{P}_{A\mbox{-}N}$
Any link on an authorized forwarding path is referred to as an authorized link. 
\sys guarantees that all the ASes on $\mathcal{P}_{A\mbox{-}N}$ can filter the traffic $T$ if $T$ traverses any unauthorized link, and all of the ASes that are not on $\mathcal{P}_{A\mbox{-}N}$ will discard traffic $T$. 

\subsection{Building Blocks} 
\label{subsec:BuildingBlocks}
\parab{Forwarding Commitment} \sys enhances the security of inter-domain routing and forwarding by building a publicly verifiable view on \emph{forwarding commitments}. At a high-level, a routing commitment (\fcshort) of an AS is a cryptographically-signed primitive that binds the AS's routing decisions (\eg willing to forward traffic for a prefix via a certain two-hop pathlet). With this view, ASes are able to \first evaluate the authenticity (or security) of the control plane BGP announcements based on committed routing decisions from relevant ASes, and \second ensure that the data plane forwarding is consistent with the routing decisions committed in the control plane.

\parab{BGP Path Validation} Consider an illustrative example using the four-AS topology shown in Figure~\ref{example}. 
Suppose AS $C$ receives a BGP announcement $\mathcal{P}_A{:}A {\leftarrow} B {\leftarrow} C$ from its neighbor $B$. If AS $C$ decides to further advertise this path to its neighbor $D$ based on its routing policy, it generates a \fcshort $\mathcal{F}_{\{B,C,D,\mathcal{P}\}}$, adds it to the path attribute field of the BGP announcement, and forwards the BGP update message to $D$. 
When AS $D$ receives the route from $C$, it can determine the authenticity of the current AS path by verifying the list of \fcs carried in the BGP announcement correctly reflects the AS path. The exact computation of $\mathcal{F}_{\{B,C,D,\mathcal{P}\}}$ is explained in \S~\ref{subsec:fc_design}. 

\parab{Forwarding Validation} To enable forwarding validation, ASes need to announce the traffic-\fcs binding relationship. Specifically, suppose AS $D$ confirms that the AS-path $C{\rightarrow}B{\rightarrow}A$ reaching $\mathcal{P}_A$ is legitimate, it binds the traffic $\langle \textsf{src}{:}\mathcal{P}_D, \textsf{dst}{:}\mathcal{P} \rangle$ (where $\mathcal{P}_D$ is the source prefix owned by AS $D$) with the \fcshort list $\{ \mathcal{F}_{\{B,C,D,\mathcal{P}\}};\mathcal{F}_{\{A,B,C,\mathcal{P}\}};\mathcal{F}_{\{null,A,B,\mathcal{P}\}} \}$, and then publicly announces the binding relationship. 
Upon receiving the relationship, other ASes can build traffic filtering rules to enable path validation. For instance, by interpreting the binding relationship produced by AS $D$, AS $C$ confirms that the traffic $\langle \textsf{src}{:} \mathcal{P}_D, \textsf{dst}{:}\mathcal{P}_A \rangle$ shall be forwarded on link $L_{CD}$, and AS $B$ confirms that the traffic shall be forwarded on link $L_{BC}$. 
Network traffic violating these binding rules are not legitimate. 

To enable network-wide forwarding verification, these binding rules are broadcast globally (instead of just informing the ASes on the AS-path) so that off-path ASes can also discard unauthorized flows. We design a lightweight protocol, optimized using BGP semantics, to achieve fast binding relationship synchronization globally. 
\section{Control Plane Validation}

In this section, we discuss the specific design of $FC$s in \sys and how to generate, transfer $FC$s and eventually complete path validation for the BGP announcement. Based on this, we discuss the impact of partial deployment scenarios on the validation mechanism. For security analysis of BGP validation, please see \S~\ref{sec:analysis}.

\subsection{Forwarding Commitment}\label{subsec:fc_design}

We need to address three concerns in designing forwarding commitments (\fcs): \first ensuring that \fcs are not maliciously spliced to forge announcement paths, \second minimizing duplicate validation of \fcs during route changes, and \third minimizing the propagation range of \fcs to protect the privacy of routing policies.

The \fcs in \sys are 
\emph{path-independent}. We call a tuple $\langle previous AS, current AS, next AS \rangle$ on an AS path a \emph{pathlet}.\fcshort is designed to verify the forwarding intent on a pathlet. In particular, suppose that AS $C$ receives a BGP update $\mathcal{P}{:}S {\leftarrow} A {\leftarrow} B$, if $B$ prefers to further advertise this path to its neighbor AS $C$, AS $B$ computes $\mathcal{F}_{\{A,B,C,\mathcal{P}\}}$ to authenticate the preference on the $B {\leftarrow} C$ hop as follows:
\begin{equation}
    \mathcal{F}_{\{A,B,C,\mathcal{P}\}} = \Big\{\mathcal{H}(A,B,C, \mathcal{P})_{\textsf{Sig}_B} ~||~ \textsf{A}  ~||~ \textsf{B}  ~||~ \textsf{C} \Big\},
\end{equation}
where $\mathcal{H}$ is a (public) secure one-way hash function, $\textsf{A}$ is the previous AS, $\textsf{B}$ and $\textsf{C}$ are endpoints of this hop, $\textsf{Sig}_B$ is the signature using private key of AS $B$.


\parab{Resilience Analysis} Because an \fcshort specifies a per-pathlet intent of the signing AS. The adversary is motivated to strategically splice these \fcs to construct a nonexistent path that could be authenticated by these strategically combined \fcs. However, the structure of the \fcshort guarantees that the adversary cannot forge a valid AS path when \sys is universally deployed. We provide a rigorous security analysis in \S\ref{sub:SecurityofCP}.

\parab{Minimal Validation} Binding a \fcshort to a specific pathlet instead of the entire BGP path ensures that ASes will not redundantly generate \fcs in case of partial routing updates. For instance, if a BGP path changes from $\mathcal{P}{:} A {\leftarrow} B {\leftarrow} C {\leftarrow} D {\leftarrow} E$ to $\mathcal{P}{:} A {\leftarrow} F {\leftarrow} C {\leftarrow} D {\leftarrow} E$, $D$ does not need to repeatedly certify the same routing decision of using the pathlet $C {\leftarrow} D {\leftarrow} E$ to carry traffic $\langle \textsf{src}{:}E, \textsf{dst}{:}A\rangle$. As quantified in \S~\ref{subsubsection:Internet Validation}, this greatly reduces the aggregate validation overhead in a dynamic network like the Internet. 

\parab{Privacy of Routing Policy} \fcshort contains the information to reflect the routing policy of an AS. Therefore, it is necessary to consider avoiding the leakage of the routing policy of an AS due to \fcshort. To achieve this, a new Path Attribute is used to deliver \fcs in BGP update messages (see \S~\ref{p-fc}.), similar to BGPsec. Intercepting the \fcshort requires intercepting the corresponding BGP update message first. The content recorded in the \fcshort does not exceed the information carried in the BGP update, and the propagation scope is consistent with the BGP update. Therefore, in the existing inter-domain routing system, the AS does not cause additional risk of routing policy leakage due to \fcshort.

\parab{Propagation} We use a new type of path attribute in the BGP update message to carry the FC, limited by space, please see ~\ref{p-fc} in the Appendix for the detailed message format.

\subsection{BGP Path Validation}\label{subsec:bp_confirm}
\begin{figure}[t]   
\centering
\includegraphics[width=0.45\textwidth]{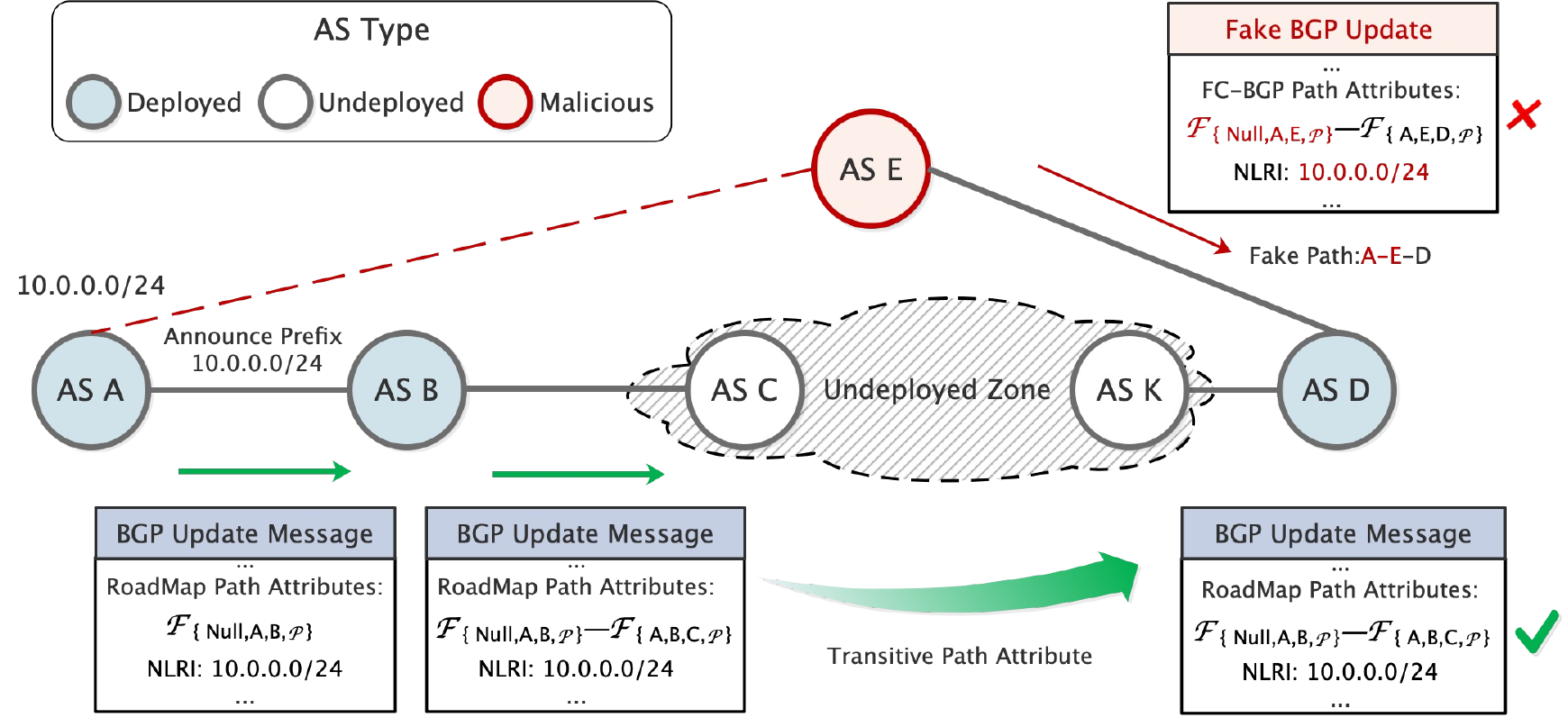}
\caption{Example of BGP Path Validation}
\label{Bgppath}
\end{figure}

In this segment, we describe the concrete path validation protocol based on \fcs. 
Consider the topology in Figure~\ref{Bgppath}. Suppose that $A$ owns the prefix $\mathcal{P}$ $10.0.0.0/24$ (which can be verified by all other ASes via RoV). It  commits and announces the route for this prefix to $B$, by computing a \fcshort $\mathcal{F}_{\{Null,A,B,\mathcal{P}\}}$ and then populating the BGP update message with the proper \fcshort path attribute. 

\parab{Partial Deployment} 
Given the scale and heterogeneity of ASes, it is unlikely to witness a simultaneous global upgrade. Therefore, we first discuss the partial deployment case where only part of the ASes are upgraded to support \sys. 
Given the topology shown in Figure~\ref{Bgppath}, the BGP announcement sent by $B$ needs to transit through a legacy region before reaching the next AS $D$ that supports \sys. Consider AS $C$ as an example in the legacy region. Upon receiving the BGP update message from $B$, AS $C$ receives and stores it in the \textsf{Adj-RIB-In} table and starts normal BGP processing. 
Based on the setting of \textsf{Attr.TYPE}, AS $C$ will continue to pass the \fcs path attribute even if $C$ does not recognize it. 
Afterward, this update message is pushed into the \textsf{Adj-RIB-Out}, after being processed by the local routing policy engine of $C$, and waits to be sent to the next hop. 
Therefore, \sys requires no extra support from legacy ASes to propagate \fcs in the network, the key to achieving incremental deployability. In partial deployments, successive AS deployments from the source can guarantee a secure sub-path. The longer the security sub-path is, the closer the malicious AS must be to the target AS for traffic hijacking. Please see \S~\ref{sub:SecurityofCP} for the security benefits of partial deployment.


\parab{Route Selection} 
ASes determine the authenticity of a BGP update by validating that \fcs carried in the path attribute are consistent with the AS-path, \ie there exists a correctly signed \fcshort to certify each pathlet on the AS-path. 
Considering the partial deployment, there are four types of paths. 
\begin{itemize}[leftmargin=*]
    \item Trusted path: every pathlet in the AS-path is certified by a properly signed \fcshort. 
    \item Partially trusted path: all carried \fcs are consecutive, correct, and starting from the source AS, but at least one pathlet exists (\ie not certified by any \fcshort). 
    \item Legacy path: no \fcs are carried in the update, \ie all prior ASes on the AS-path are legacy. 
    \item Suspicious path: at least one upgraded hop on the AS-path misses a corresponding \fcshort. For instance, as shown in Figure~\ref{Bgppath}, suppose AS $D$ receives a BGP advertisement for path $\mathcal{P}{:}A{\leftarrow}E{\leftarrow}D$ from the malicious AS $E$, AS $D$ expects a valid \fcshort $\mathcal{F}_{\{Null,A,E,\mathcal{P}\}}$ to certify the hop $A{\leftarrow}E$ (because AS $A$ is not legacy). Since AS $E$ cannot produce a correctly signed $\mathcal{F}_{\{Null,A,E,\mathcal{P}\}}$, this path $\mathcal{P}{:}A{\leftarrow}E{\leftarrow}D$ is suspicious. 
\end{itemize}

To favor secure routing, the ASes joining the \sys ecosystem shall design routing policies in their local routing policy engines, preferring trusted paths over partially trusted paths over legacy paths over suspicious paths.



\parab{Update and Withdraw} 
Similar to BGPsec, \sys is well compatible with the route update and withdrawal mechanism of BGP. An AS $K$ supporting \sys needs to store the \fcshort list corresponding to the validated AS-Path. When a new BGP peer appears, $K$ needs to announce its own route selection along with the corresponding \fcshort list to the new neighbor. When a route update arrives, AS $K$ performs route selection according to the local routing policy and updates the storage of corresponding \fcshort list. If a route withdrawal occurs due to a link disconnection or a change in the peer relationship, etc., the AS $K$ would delete the \fcshort list corresponding to the withdrawn route when it receives the withdrawal message, and at the same time continue to announce this withdrawal message to other neighbors.

\section{Data Plane Validation}

In this section, we discuss the mechanism design of \sys in the data plane. The declaration of legitimate forwarding paths and the generation of filtering information are described in \S~\ref{subsec:forward_binding}, and the collaborative filtering of off-path nodes is described in \S~\ref{subsec:offpath}. In \S~\ref{subsec:bing_sync}, we will present a novel protocol design for guaranteeing global consistency of the information used to filter forged traffic.

\begin{figure}[t]    
\centering
\includegraphics[width=0.45\textwidth]{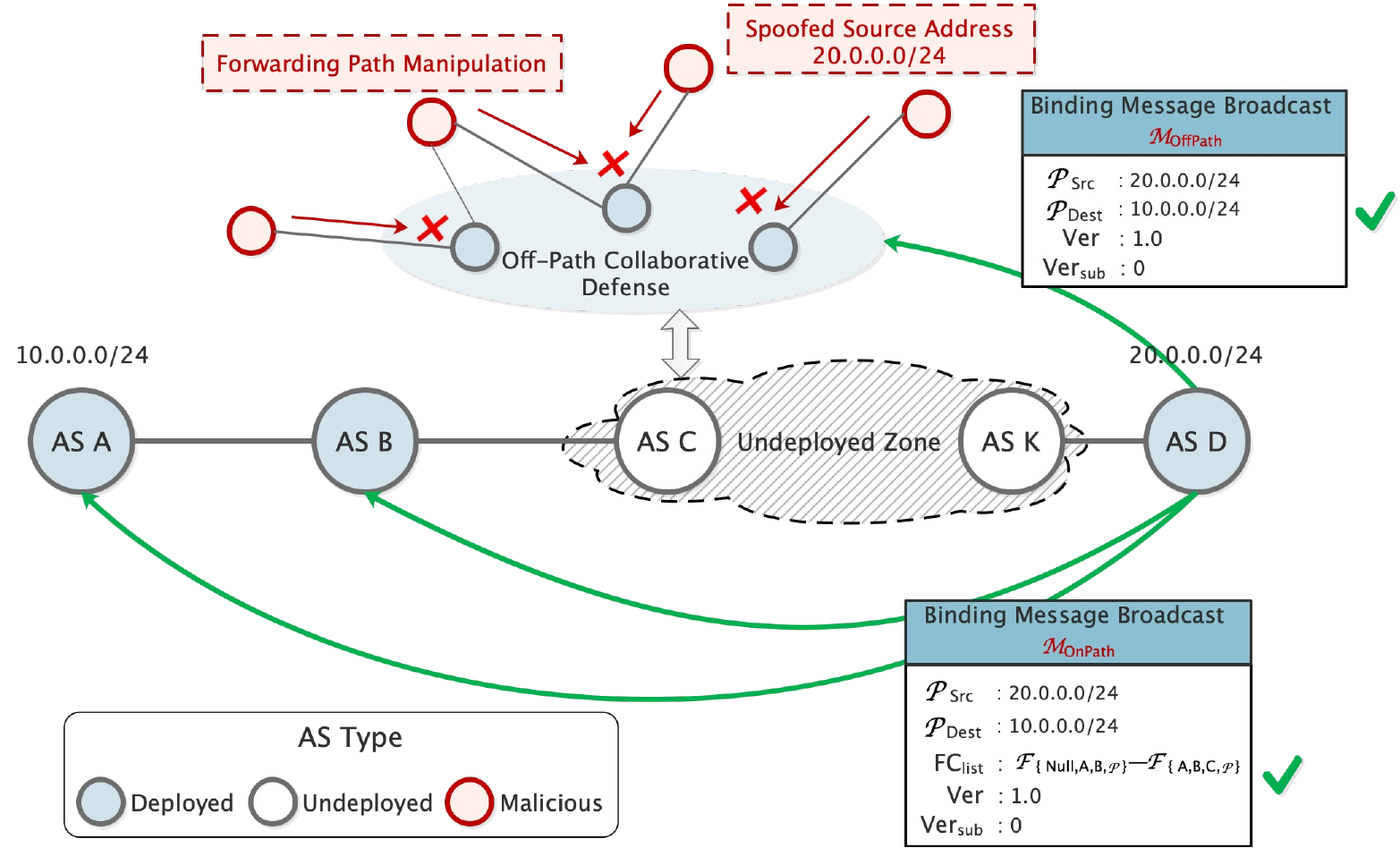}
\caption{Example of Forwarding Validation}
\label{Forwardingpath}
\end{figure}



\subsection{Forward Binding}
\label{subsec:forward_binding}
Although \fcs are initially invented to validate control plane BGP announcements, our key insight into the applicability of \fcs in data plane is that the correctness of a forwarding path can be directly certified based on the \fcshort lists carried in BGP updates. 
Specifically, in the topology shown in Figure~\ref{Bgppath}, once AS $D$ selects the BGP path $\mathcal{P}{:}A{\leftarrow}B{\leftarrow}C{\leftarrow}K{\leftarrow}D$, it implies to all on-path and upgraded ASes (\ie $A$ and $B$) that the network traffic $\langle \textsf{src:}\mathcal{P}_D,\textsf{dst:}\mathcal{P}\rangle$ ($\mathcal{P}_D$ represents the prefix owned by $D$) shall take the forwarding path $D{\rightarrow}K{\rightarrow}C{\rightarrow}B{\rightarrow}A$. 
To make this implicated forwarding path explicit, AS $D$ can send the \fcs list (\ie $\{\mathcal{F}_{\{A,B,C,\mathcal{P}\}};\mathcal{F}_{\{Null,A,B,\mathcal{P}\}} \}$) to AS $A$ and $B$, as shown in Figure~\ref{Forwardingpath}. Afterwards, AS $B$ confirms that the traffic $\langle \textsf{src:}\mathcal{P}_D,\textsf{dst:}\mathcal{P}\rangle$ must inbound using the hop $C{\rightarrow}B$, and AS $A$ confirms the traffic must inbound via the the hop $B{\rightarrow}A$. 

We define the above process of certifying the entire forwarding path for specific network traffic (identified by the source-prefix and destination-prefix pair) as \emph{forward binding}. Formally, the binding message $\mathcal{M}_{\textsf{OnPath}}$ is given in Equation~(\ref{eqn:binding_onpath}). 

\begin{equation}
   \label{eqn:binding_onpath}
   \mathcal{M}_{\textsf{OnPath}} = \Big\{\mathcal{P}_{\textsf{src}},\mathcal{P}_{\textsf{dst}},\textsf{FC}_{\textsf{list}}, \textsf{Ver}, \textsf{Ver}_{\textsf{sub}} \Big\}_{\textsf{Sig}_\textsf{src}},
\end{equation}
where $\mathcal{P}_{\textsf{src}}$ and $\mathcal{P}_{\textsf{dst}}$ represent source and destination prefixes, respectively. \textsf{Ver} and $\textsf{Ver}_{\textsf{sub}}$ are version numbers required in the synchronizing binding message (see details in \S~\ref{subsec:bing_sync}). 

Upon receiving  $\mathcal{M}_{\textsf{OnPath}}$ binding, on-path ASes verify message and then  install local data plane filters to discard the traffic forwarded over illegitimate/uncertified paths. The verification includes \first $\mathcal{M}_{\textsf{OnPath}}$ is correctly signed; \second the issuer of $\mathcal{M}_{\textsf{OnPath}}$ is the correct owner of prefix $\mathcal{P}_\textsf{src}$, and \third $\textsf{FC}_{\textsf{list}}$ does include a \fcshort signed by the on-path AS. 

\subsection{Off-Path Collaborative Filtering}
\label{subsec:offpath}
The forward binding can be extended to off-path ASes to enable large-scale traffic filtering. In particular, in Figure~\ref{Forwardingpath}, AS $D$ can broadcast the binding relationship for traffic $\langle \textsf{src:}\mathcal{P}_D,\textsf{dst:}\mathcal{P}\rangle$ to off-path ASes. 
The key difference between the on-path binding message and off-path binding message is that latter does not specify any \fcshort list, as shown in Equation~(\ref{eqn:binding_offpath}). As a result, all off-path ASes only learn that they should not expect to serve traffic $\langle \textsf{src:}\mathcal{P}_D,\textsf{dst:}\mathcal{P}\rangle$, instead of learning the actual forwarding path for the traffic. 

\begin{equation}
\label{eqn:binding_offpath}
   \mathcal{M}_{\textsf{OffPath}} = \Big\{\mathcal{P}_{\textsf{src}},\mathcal{P}_{\textsf{dst}}, \textsf{Ver}, \textsf{Ver}_{\textsf{sub}} \Big\}_{\textsf{Sig}_\textsf{src}}.
\end{equation}

The off-path collaborative filtering in \sys is particularly effective to filter network-wide illegitimate traffic. As shown in Figure~\ref{Forwardingpath}, off-path ASes may receive traffic $\langle \textsf{src:}\mathcal{P}_D,\textsf{dst:}\mathcal{P}\rangle$ due to either source spoofing or path manipulation. In both cases, they can discard these flows based on the binding message received from AS $D$. 
The key advantage of this off-path collaborative filtering is that it is completely location-independent, \ie the off-path filtering is effective no matter how far is it between the filtering AS and the actual on-path ASes, and no matter how many legacy regions are between them. 
In \S~\ref{sub:SecurityofDP}, we show that with only 0.7\% upgraded ASes, \sys can filter 50\% of all possible forged traffic in Internet. 

\subsection{Binding Message Synchronization}
\label{subsec:bing_sync}
The key to enable forwarding validation is to globally publish the forward binding messages. 
Intuitively, \sys can leverage Resource Public Key Infrastructure (RPKI) extension to publish these information. However, the synchronization cycle of RPKI is too long: the relying parties pull updates from  publication points every refresh interval, which range between 10 minutes and 1 hour, depending on the relying party implementation~\cite{hlavacek2022stalloris,kristoff2020measuring}. 

Alternatively, we can resort to consensus protocols. Existing BFT applications require all transactions to be ordered and executed sequentially, the consensus protocols must use some complex sub-protocols to meet the requirements. For example, asynchronous BFTs~\cite{gao2022dumbo,duan2018beat,yang2022dispersedledger} rely on the complex multi-value byzantine agreement (MVBA) sub-protocol~\cite{cachin2001secure,abraham2019asymptotically,lu2020dumbo} for ordering the transactions, which incurs significant consensus in WAN. (Partially) synchronous BFT ~\cite{castro1999practical,clement2009making,yin2019hotstuff} relies on a leader to order the transactions and thus can achieve lower consensus latency, but they must use the view-change sub-protocol to handle leader failures. They also assume bounded communication latency which is not guaranteed to be correct in the Internet. In addition, as we will demonstrate in \S~\ref{subsubsec:eval:sync_eff}, directly applying (partially) synchronous BFT protocols to disseminate binding messages will easily overflow the leader's network bandwidth.


\parab{Protocol Overview}
In this segment, we present the protocol specifically designed for synchronizing binding messages in the Internet. 
The key design to achieve fast and secure binding message publication is the decoupling of broadcast and consistency check. 
Specifically, after confirming a path, an AS broadcasts the corresponding binding message (\ie $\mathcal{M}_{\textsf{OnPath}}$ and $\mathcal{M}_{\textsf{OffPath}}$) to all on-path and off-path ASes. 
Upon receiving valid binding messages, these ASes can immediately install packet filter rules \emph{without waiting for any form of further confirmation}. 
This is because all binding messages are self-proving: \ie ASes cannot generate false binding messages on behalf of other ASes. Therefore, all received (valid) binding messages are effective immediately. 
Afterwards, a consistency check is \emph{executed periodically} to ensure that the correct ASes have the same view on binding relationships for achieving off-path collaborative filtering. 
This protocol does not require any transmissions of the actual binding messages, which significantly reduces validation overhead. We quantify this benefit in \S~\ref{subsubsec:eval:sync_eff}. 


\subsubsection{Binding Version Consistency Check Protocol}\label{subsubsec:version_view}
\parab{Binding Version View}
Suppose there are $K$ ASes in the current ecosystem of \sys. Each of them maintains a $K$-dimensional vector named \emph{Binding Version View, \textsf{BVV}}, where the $i$-th element records the version number of the latest binding message received from AS $i$. Specifically, the binding version view of AS $j$ is as follow:

\begin{equation}
   \mathcal{V}_{j} = \Big\{\textsf{(ASN}_{i},\textsf{Ver}_{i}{)} ~|~ \textsf{i}\in \textsf{(1,K)}  \Big\},
\end{equation}
where $\textsf{ASN}_{i}$ is the AS Number of AS $i$, and $\textsf{Ver}_{i}$ is latest version number from AS $i$. 

\parab{Periodical Consistency Check} In each period, one AS (\ie the round-leader) is responsible for advertising its local BVV to all other members via a Reliable Broadcast Communication (RBC) protocol ~\cite{bracha1987asynchronous}. 
For instance, in the $v$-th period, AS $L$ (where $L = v \pmod{N}$) is selected. 
A secure RBC requires at least $n - \lfloor n/3 \rfloor$ ASes are honest and IP reachable to each other.  
In \S~\ref{subsubsec:sharding}, we show that it is possible to relax the global connectivity requirement via sharding. 

Upon receiving the BVV (denoted as $\mathcal{V}^{g}$) from the $RBC_L$ (the RBC protocol initiated by AS $L$), an AS $j$ compares $\mathcal{V}^{g}$ with its local BVV ($\mathcal{V}_{j}$) as follows:





1) If the $\textsf{Ver}_{i}$ of AS $i$ in $\mathcal{V}^g$ is greater than $\textsf{Ver}_{i}$ in $\mathcal{V}_{j}$, then AS $j$ misses at least one new binding message from AS $i$. In this case, besides simply adopting a newer version, AS $j$ can proactively request missed binding messages from AS $L$. The rationale is that the missing of prior updates is an indicator of weak or unstable network connectivities for AS $j$, and therefore the proactive requests are helpful.

2) If $\textsf{Ver}_{i}$ in $\mathcal{V}^g$ is the same as the local copy, indicating view consistency among AS $i$, $j$, and $L$. 

3) If $\textsf{Ver}_{i}$ in $\mathcal{V}^g$ is smaller than the local version, then AS $L$ misses some updates from AS $i$. In this case, AS $j$ should send the more recent $\textsf{Ver}_{i}$ to AS $L$.

Considering that the round-leader AS in the $v{-}1$-th period may be faulty and refuse to start the RBC, the round-leader AS in the period $v$-th can initiate its own RBC without waiting for the finish of prior RBCs. This will not undermine protocol correctness, as shown in \S~\ref{subsubsec:correctness}.

\subsubsection{Path Update and Withdrawal}



\begin{figure}[t]    
\centering
\includegraphics[width=0.45\textwidth]{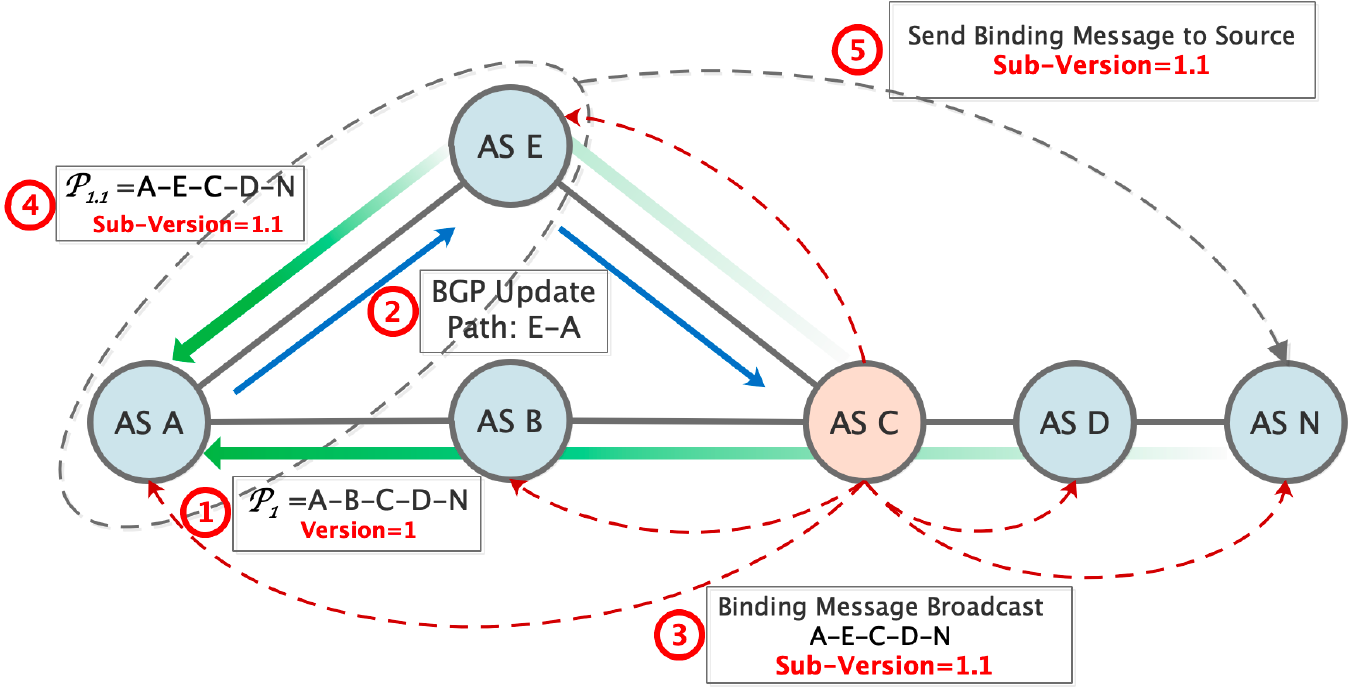}
\caption{Example of sub-version update}
\label{subversion-part1}
\end{figure}

When the BGP path is partially updated on the control plane, it takes time to propagate the new binding message that reflects the path change. We thus design a subversion-based protocol to avoid incorrect filtering. 
Take the Figure~\ref{subversion-part1} as an example. The source AS $N$ originally binds its traffic to AS $A$ to the path $\mathcal{P}{_1}{:} A {\leftarrow} B {\leftarrow} C {\leftarrow} D {\leftarrow} N$. Later, AS $C$ learns a new path to AS A $\mathcal{P}{1.1}{:} A {\leftarrow} E {\leftarrow} C$, and according to AS $C$'s routing policy, it chooses AS $E$ as the next hop to reach $A$. In this case, AS $C$ will generate a new binding message, where the $\textsf{Ver}$ is the version number in the original binding message of path $\mathcal{P}_{1}$, and $\textsf{Ver}_\textsf{sub}$ is a sub-version number added by AS $C$ to indicate a partial path update. 
After that, AS $C$ sends the binding message with the subversion number to other ASes. Meanwhile, AS $E$ and AS $A$, as the on-path nodes, also forward the binding message generated by AS $C$ to the source AS $N$. 
When the new BGP update eventually reaches the AS $N$, it validates the subversioned binding message generated by AS $C$ against the BGP update. If the verification passes, AS $N$ updates the master version number and generates a new binding message. Otherwise, AS $N$ can also overwrite the subversioned binding message via a higher master version number. 

When a path is withdrawn and there is no other optional path, an AS can  generate a withdrawal binding message with $\textsf{Ver}{=-1}$ for the corresponding traffic. Upon receiving valid withdrawal  messages, ASes should remove the corresponding filters.

\subsubsection{Sharding}
\label{subsubsec:sharding}
To honor the distributed governance of the Internet, and meanwhile reducing overhead for consistency check, 
it is possible for the ASes in \sys to form several regions (possibly according to their geographical location and/or business relationships), where each region operates independently on synchronizing the BVVs within the region. 
Afterwards, the round-leader AS is responsible for broadcasting the post-check BVVs to all other regional ASes. Note that the instant broadcast of binding messages by individual ASes can still be global without considering the boundary of regions.


During the course of \sys deployment, ASes may subscribe or leave the ecosystem of \sys. Thus, the network needs a public view on the list of active members. We can adopt the membership reconfiguration mechanisms in~\cite{castro2002practical,duan2022foundations}. 
\vspace{-0.8cm}
\subsubsection{Correctness Analysis}
\label{subsubsec:correctness}

In this segment, we prove that all correct ASes will eventually converge to a consistent view on the binding messages. Towards this end, we assume that the network connectivity within a region is rich enough to satisfy  that \first given a region with $n$ ASes, at least $n-\lfloor n/3 \rfloor$ ASes are honest and are directly IP reachable to each other; \second all the correct ASes are \emph{eventually IP reachable} (\ie for any two correct ASes in a region, there exists at least one \emph{broadcast chain} between them such that a message sent by one AS will eventually reach the other AS after being relayed/broadcasted multiple times by the intermediate ASes in the region). 
The assumption for the network connectivities across regions is relaxed: we only assume that any two honest ASes across different regions are eventually IP reachable, \ie they do not need to be directly IP reachable. This honors the fact that a strong adversary controlling multiple ASes could block the direct connectivities between regions~\cite{2020A}. 


\parab{Intra-Region Correctness} 
In the period $v$, the BVV sent by the round-leader AS (denoted as AS $L$) will reach all other correct ASes that are reachable with AS $L$ (assured by the correctness of the underlying RBC protocol). 
In a corner case where an AS $J$ that is neither IP reachable to AS $L$ nor IP reachable to all the ASes that are reachable to AS $L$, AS $J$ cannot receive the BVV in this round. 
However, AS $J$ will eventually get the BVV as follows. Suppose that AS $L$ and AS $J$ are eventually IP reachable via the following sequence of intermediate ASes: $\mathcal{P}{:} AS_L {\rightarrow} AS_{k_1} {\rightarrow} AS_{k_n} {\rightarrow} AS_{J}$ where any two adjacent ASes in the sequence are directly IP reachable. In this round, both $AS_{k_1}$ and $AS_{k_n}$ can receive the BVV. When  $AS_{k_n}$ becomes the round-leader in later round, $AS_{J}$ will receive BVV, based on which it can synchronize any previously missed binding messages.


\parab{Inter-Region Correctness} 
The correctness of inter-region synchronization is similarly held because the corrected AS across regions are eventually IP reachable. Yet, synchronization across regions may takes more rounds of broadcast. 
In \S~\ref{subsubsec:eval:sync_eff}, we evaluate the latency of running our synchronization protocol over a large overlay network with 100 ASes distributed across the globe. 

\section{Security Analysis}
\label{sec:analysis}

In this section, we analyze the design of \sys, discuss its security, deployability and analyze the security benefits of deploying \sys on the Internet. 

\subsection{Security of Control Plane}
\label{sub:SecurityofCP}

\subsubsection{Complete Deployment} 
\begin{figure}[t]  
\centering
\includegraphics[width=0.45\textwidth]{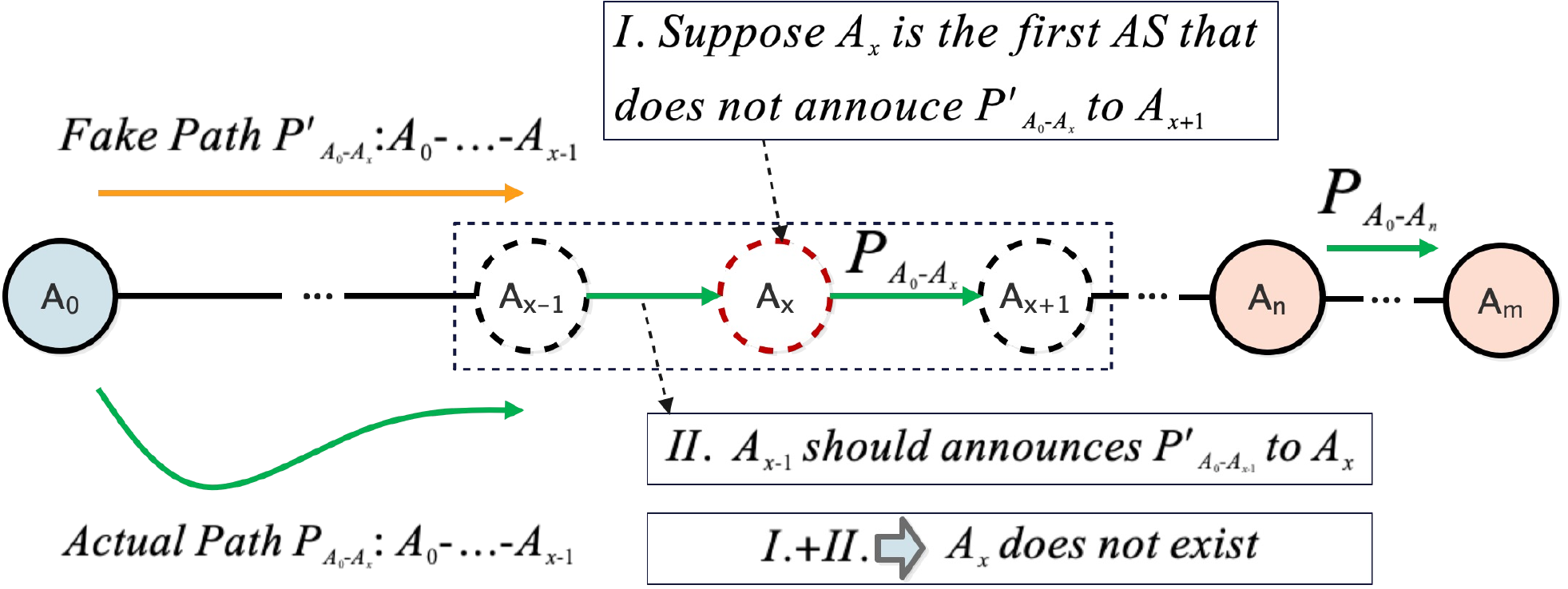}
\caption{Security analysis in case of complete deployment. Since each \fcshort binds a pathlet, any AS $A_{x}$ on the fake path should announce the path learned from AS $A_{x-1}$ to AS $A_{x+1}$, starting from the actual pathlet $A_{0}\mbox{-}A_{1}\mbox{-}A_{2}$. Therefore AS $A_n$ can not piece together a non-existent path.}
\label{CompleteDeployment}
\end{figure}

In this section, we demonstrate that the adversary cannot forge a valid AS path when \sys is universally deployed.
We prove the above statement using the example shown in Figure~\ref{CompleteDeployment}.   
Without loss of generality, 
suppose that AS $A_{n}$ is the first compromised AS on the AS-path of a BGP update message originated from AS $A_{0}$. 
The adversary cannot forge legitimate \fcs signed by other ASes, but it can collect the \fcs carried in legitimate BGP update messages. Thus, the adversary tries to build a seemingly valid path by strategically combining these collected \fcs such that it can find a valid \fcshort for every individual hop on the path. 

We use the following definitions for convenience.
\begin{itemize}[leftmargin=*]
\item \emph{Actual path} $\mathcal{P}_{i\text{-}j}$ represents the announced AS-path from AS $i$ to $j$ that can be authenticated with \fcs. 

\item \emph{Fake path} $\mathcal{P}^{\prime}_{i\mbox{-}j}$ represents any non-existent path that the adversary can construct by strategically combining collected \fcs. For instance, we use $\mathcal{P}^{\prime}_{A_{0}\mbox{-}A_{n}}:A_{0}\mbox{-}A_{1}\mbox{-}...\mbox{-}A_{n-1}\mbox{-}A_{n}$ to denote any fake path that the adversary can construct from AS $A_{0}$ to AS $A_{n}$. 


\item \emph{Divergence point}. The two paths $\mathcal{P}_{A_{0}\mbox{-}A_{x}}$ and $\mathcal{P}^{\prime}_{A_{0}\mbox{-}A_{x}}$ share the same origin AS. Suppose that $A_{x}$ is the first AS at which the two paths $\mathcal{P}_{A_{0}\text{-}A_n}$ and $\mathcal{P}^\prime_{A_{0}\text{-}A_n}$ diverge, meaning that in reality that $A_{x}$ announces a path $\mathcal{P}''$ that is different from $\mathcal{P}^\prime_{A_{0}\text{-}A_{x}}$ to the next AS $A_{x+1}$. We refer to $A_{x}$ as a divergence point. 

\end{itemize}

Suppose $A_{x}$ is the first divergence point. This implies two facts for AS $A_{x-1}$: \first $\mathcal{P}_{A_{0}\text{-}A_{x-1}} = \mathcal{P}^\prime_{A_{0}\text{-}A_{x-1}}$ and \second $A_{x-1}$ has announced $\mathcal{P}_{A_{0}\text{-}A_{x-1}}$ (or $\mathcal{P}^\prime_{A_{0}\text{-}A_{x-1}}$) to $A_x$. 
By assumption, the fake path $\mathcal{P}^\prime_{A_{0}\text{-}A_{n}}$ is authenticated, then the \fcshort corresponding to each two-hop pathlet on the path exists. Consider the \fcshort $\mathcal{F}_{\{A_{x-1},A_{x},A_{x+1},\mathcal{P}\}}$ that certifies the pathlet $A_{x-1}\mbox{-}A_{x}\mbox{-}A_{x+1}$ on $\mathcal{P}^\prime_{A_{0}\text{-}A_n}$. This \fcshort implies two facts: \first $A_x$ has committed to select the BGP announcement from $A_{x-1}$, which is $\mathcal{P}_{A_{0}\text{-}A_{x-1}}$; \second $A_x$ shall extend the path as $\mathcal{P}_{A_{0}\text{-}A_{x-1}} + A_x$ (which is $\mathcal{P}_{A_{0}\text{-}A_{x}}$) and announce it to the AS $A_{x+1}$. 
We have reached a contradiction. By definition of the divergence point, AS $A_x$ should announce a path $\mathcal{P}''$ to AS $A_{x+1}$ that is different from $\mathcal{P}^\prime_{A_{0}\text{-}A_{x}}$. However, we have just showed that $A_x$ should announce $\mathcal{P}_{A_{0}\text{-}A_{x}}$ to AS $A_{x+1}$. This is a contradiction because $\mathcal{P}^\prime_{A_{0}\text{-}A_{x}}$ is the same as $\mathcal{P}_{A_{0}\text{-}A_{x}}$. 
Therefore, there can be no divergence point, which indicates that the fake path $\mathcal{P}^\prime_{A_{0}\text{-}A_n}$ is essentially identical to the actual path $\mathcal{P}_{A_{0}\text{-}A_n}$.  

Consider an additional compromised AS $A_{m}$ located after $A_{n}$. As long as they do not collude with each other, it is not difficult to prove that $A_{n}$ must announce path $\mathcal{P}_{A_{0}\mbox{-}A_{n}}$ to its next neighbor $A_{n+1}$ (which could be $A_{m}$), in order to ensure that the every pathlet on the fake path can be verified via \fcs. Thus, $A_{n}$ essentially behaves in the same way as the origin AS $A_0$ (\ie it announces an actual path to its next neighbor). Therefore, by iteratively executing the above procedure, we can prove that the fake path remains identical to the actual path, regardless of the number of non-colluding ASes. 

\parab{Key Takeaways} the above security analysis shows that in a fully deployed \sys network, an adversary cannot falsely claim the authenticity of a non-existent path by strategically combining \fcs messages. Therefore, although \sys uses \emph{pathlet intent} for control plane authentication, it achieves the same level of security as BGPsec, which uses path-level intent. 
As we will show below and in \S~\ref{subsubsection:Internet Validation}, authenticating control plane paths via pathlet intent has higher  benefits in case of partial deployment and introduces much lower authentication overhead.

\subsubsection{Partial Deployment} 
\begin{figure}[t]  
\centering
\includegraphics[width=0.45\textwidth]{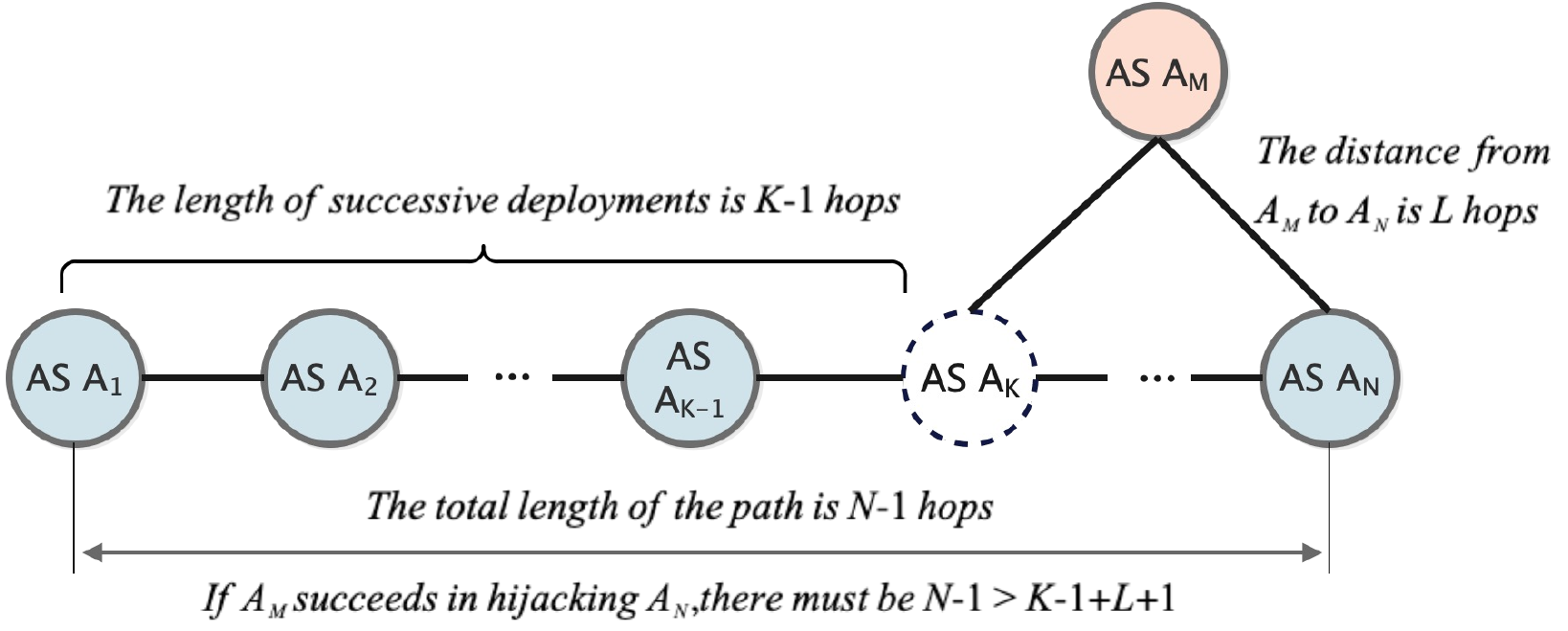}
\caption{Security Analysis for Partial Deployment.The larger $K$ is, the smaller $L$ must be. This means that the more ASes that support \sys in partial deployment scenarios, the smaller the range of possible attackers.}
\label{PartialDeployment}
\end{figure}

In this segment, we analyze the benefits of \sys in case of partial deployment. 
In Figure~\ref{PartialDeployment}, we plot a path from $A_1$ to $A_N$ with \emph{N-1} hops, where \sys is deployed successively on the ASes from $A_1$ to $A_{K-1}$ (\ie $A_{K}$ which is the first AS that does not support \sys). 
We consider the case where an upgraded AS $A_N$ located after $A_{K}$ tries to validate its receive BGP path. A compromised AS $A_M$ intends to hijack the traffic from $A_N$ to $A_1$. 
Suppose the distance between $A_M$ and $A_N$ is $L$ hops. Because $\mathcal{F}_{\{A_{K-2}, A_{K-1}, A_{K},\mathcal{P}\}}$ signed by $A_{K-1}$ exists, the best option for $A_M$ is to pretend to be a neighbor of AS $A_K$ and construct a fake path $A_1\mbox{-}A_2\mbox{-}...\mbox{-}A_K\mbox{-}A_M\mbox{-}...\mbox{-}A_N$ with length \emph{K-1+1+L} hops. $A_M$ can successfully hijack the traffic only if \emph{N-1>K-1+1+L}, which implies that $K$ has to be smaller than \emph{N-L-1}. 

\parab{Qualitative Analysis}
Given above analysis, we analyze the path hijack attack under different  deployment cases. 
The average length of inter-domain routing paths in the Internet is around $4\mbox{-}5$ hops. We take an example of a $5$-hop path $A_1\mbox{-}A_2\mbox{-}A_3\mbox{-}A_4\mbox{-}A_5\mbox{-}A_6$, where the malicious node $A_M$ tries to hijack the traffic from $A_6$ to $A_1$. When  $A_1$ and $A_6$ are upgraded (\ie \emph{K=2} and \emph{N=6}), and $A_M$'s best option is to pretend to be a neighbor of $A_2$. In this case, the hijack attack succeeds if \emph{L<N-K-1=6-2-1=3}. This means that the ASes within two hops from $A_6$ can launch hijack attacks against $A_6$. 
By applying same analysis to larger values of \emph{K}, we can conclude that \first when both $A_1$ and $A_2$ are deployed (\ie \emph{K=3}), the possible attackers are narrowed down to $A_6$'s neighbors; \second when $A_1$, $A_2$, and $A_3$ are all deployed (\ie \emph{K=4}), a hijack attack against the path becomes infeasible. 
This example qualitatively demonstrates that why \sys provides more partial deployment benefits that BGPsec. This is because BGPsec can only secure the entire path if all of the ASes on the path are upgraded. 


\parab{Quantitative Results}
We now quantify the partial deployment benefits of \sys by studying the path hijack rate in the Internet under different deployment ratios. 
We use the CAIDA September 2023 Internet BGP dataset\footnote{https://publicdata.caida.org/datasets/routing/routeviews-prefix2as/} \footnote{https://publicdata.caida.org/datasets/as-relationships/serial-1/}. We obtain the path hijack rate as follows. 
We first sort all ASes in the dataset in descending order by the number of neighbors they have. Given a deployment rate $r$, we select the top $r\%$ ASes to upgrade either \sys or BGPsec. For each BGP path in the dataset, we study whether it is possible to hijack the path from the ASes that are L hops away from the destination AS in the path. We study four values of L (namely, 1, 2, 3, and 4) and obtain the percentage of BGP paths that are hijackable. 

\begin{figure}[t]  
\centering
\includegraphics[width=0.45\textwidth]{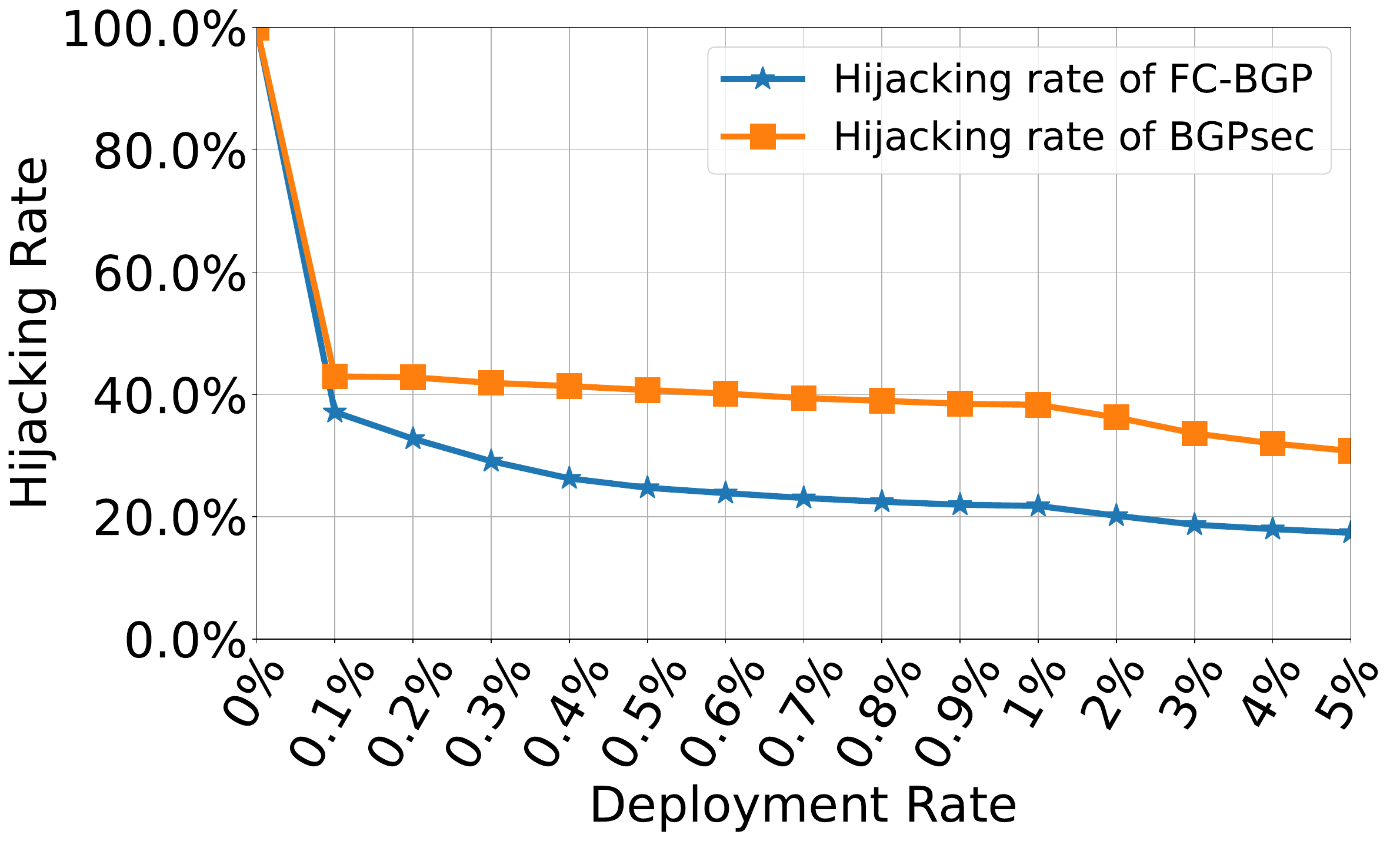}
\caption{Hijacking Rate of BGPsec and \sys for Partial Deployment.}
\label{HijackingRate}
\end{figure}

\begin{figure}[t]  
\centering
\includegraphics[width=0.45\textwidth]{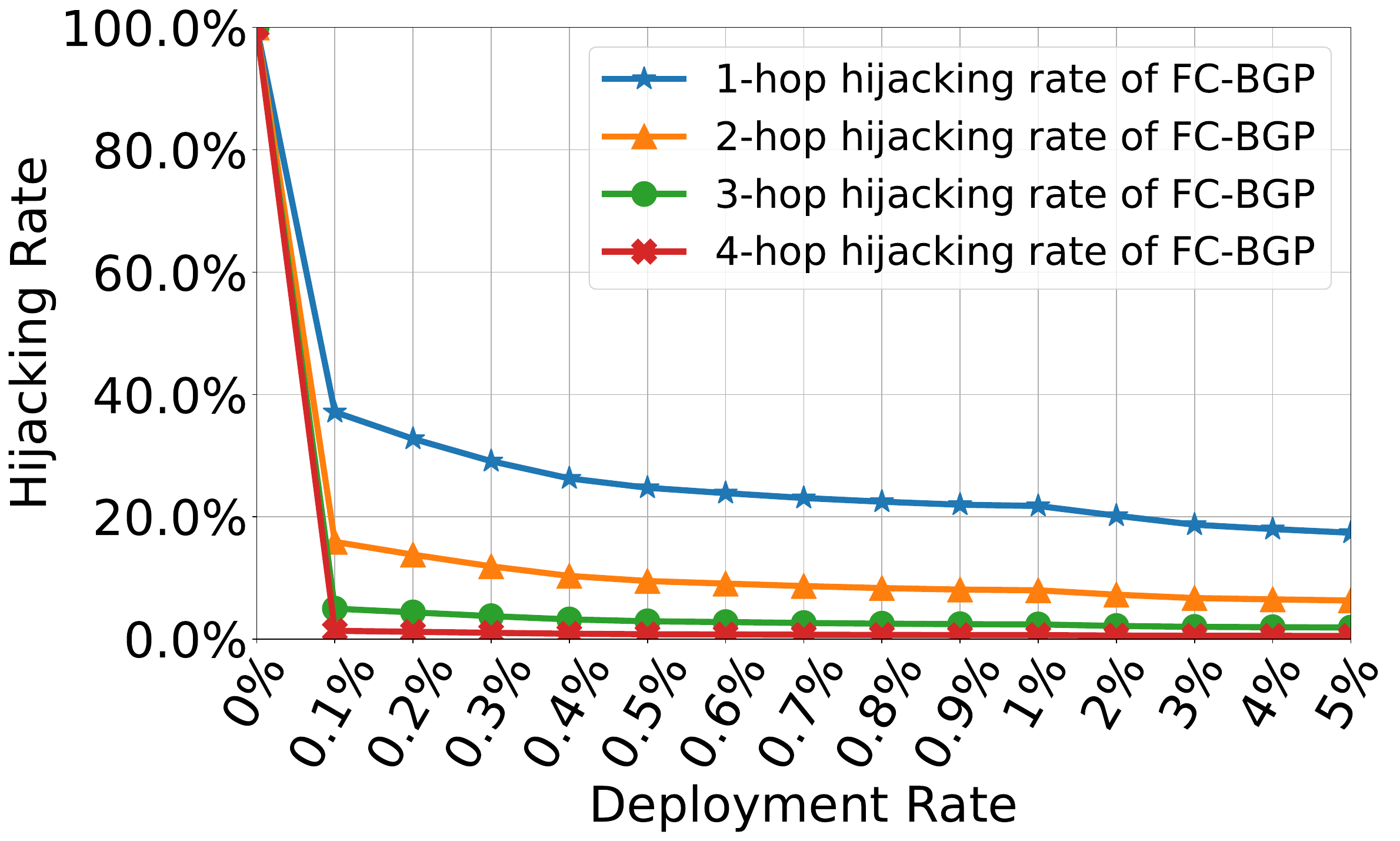}
\caption{Breakdown Hijacking Rate of \sys for Partial Deployment.}
\label{BreakdownHijackingRate}
\end{figure}

The results are plotted in Figure~\ref{HijackingRate}. The results clearly show that \sys provides strictly more benefits than BGPsec in partial deployment. For instance, at the deployment rate of 0.5\%, \sys can protect 18\% more BGP paths than BGPsec. 
We further present the breakdown of our results in Figure~\ref{BreakdownHijackingRate}. It is evident that, for a given deployment rate, the difficulty of hijacking a path increases as the attacker's distance from the destination AS increases. 
The community is standardizing proposals such as ASPA~\cite{ASPA}, which require ASes to register their direct neighbors as RPKI objects to certify the neighborhood relationship of an AS. This makes it impossible for an attacker to pretend to be a neighbor of an upgraded AS. As a result, attackers are forced to pretend to be at least two hops away from the destination AS, which reduces the probability of successful path hijacks. 


\subsection{Security of Data Plane}
\label{sub:SecurityofDP}
In this segment, we discuss the security benefits of \sys on authenticating data plane forwarding. 

\begin{figure}[t]  
\centering
\includegraphics[width=0.45\textwidth]{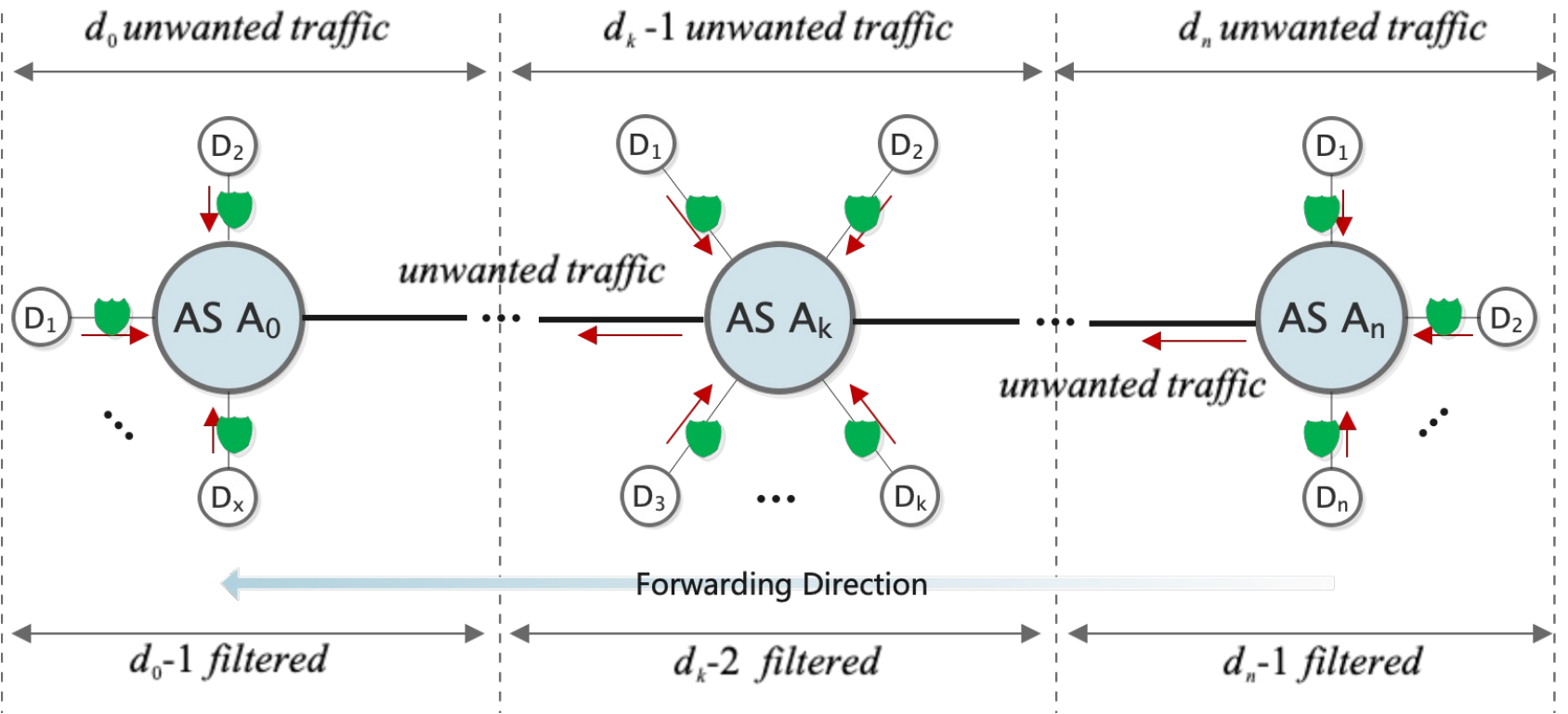}
\caption{Security Analysis for Data Plane.}
\label{DataPlane}
\end{figure}

\parab{Metric Definition}
Consider the example topology in Figure~\ref{DataPlane}, which shows a BGP path $A_{0}\mbox{-}A_{1}\mbox{-}...\mbox{-}A_{n}$. On the data plane, the network traffic is expected to flow from $A_{n}$ to $A_{0}$, in the opposite direction of the BGP path. 
Then there are three types of unwanted traffic for this forwarding path: \first traffic destinated to $A_{0}$ but sent from a source AS that is not on the BGP path (\ie from an off-path AS that has upgraded to support \sys)\footnote{We exclude the traffic is sent from a legacy AS, because the legacy AS will not propagate the binding messages.}; and \second traffic sent from an on-path AS to $A_{0}$ but taking a different forwarding path than the authorized path. This means that the traffic first diverge from the authorized path to traverse some other ASes and then re-enter the authorized path; \third traffic sent from an on-path AS to $A_{0}$ but the source AS spoofs the source address of another on-path AS. For instance, AS $A_{k}$ spoofs the source address of AS $A_{k+1}$ and sends the traffic to  AS $A_{k-1}$ over the authorized path. 

We define the filtering rate $\mathcal{F}$ as a metric to quantify the security benefits on the data plane. The filtering rate is calculated as the ratio of the amount of unwanted traffic that is successfully filtered, $N$, to the total amount of unwanted traffic that may be sent to the authorized path, $M$. 

\parab{Analysis Formulation}
Denote the number of neighbors of AS $A_{k}$ as $D_{k}$. Then the total number of ASes through which the unwanted traffic may enter the authorized path is $D_0+(D_1-1)+...+(D_{n-1}-1)+D_n=2+\sum_0^n (D_{k}-1)$. For instance, $A_{k}$ may receive the first two types of unwanted traffic from all its neighbors except for $A_{k-1}$ and the third type of unwanted traffic from itself. 
We referred to these ASes as bridges. 
For the sake of simplicity, we assume that the amount of unwanted traffic carried by each bridge is the same. Then we have $M = 2+\sum_0^n (D_{k}-1)$. 



If $A_{k}$ does not deploy \sys, it cannot filter any unwanted traffic. 
If $A_{k}$ deploys \sys, we have  

\begin{itemize}[leftmargin=*]
\setlength\itemsep{0em}
\item If \emph{0<k<n}, $A_{k}$ can filter all type-I unwanted traffic, and all type-II unwanted traffic expect for the traffic enters from $A_{k+1}$. It cannot filter type-III unwanted traffic. 
\item If \emph{k=n}, $A_{n}$ can filter all unwanted traffic expect for the type-III unwanted traffic. 

\item If \emph{k=0}, $A_{0}$ can filter all unwanted traffic expect for the type-II unwanted traffic received from $A_{1}$. 
\end{itemize}

Denote the amount of unwanted traffic that AS $A_k$ can filter as $y_{k}$. We have 

\begin{equation}
  y_{k}=\begin{cases}
  (D_{0}-1)*t_{0},& k=0;\\
  (D_{k}-2)*t_{k},& 0<k<n;\\
  D_{n}-1,& k=n.
  \end{cases}
\label{equ:filter}
\end{equation}

\begin{equation}
  t_{k}=\begin{cases}
  0,& A_{k} \text{ deploys \sys;}\\
  1,& A_{k} \text{ dose not deploy \sys.}
  \end{cases}
\label{equ:deploy}
\end{equation}

The total amount of filtered unwanted traffic is $N = \sum_0^n y_{k}$. 

\begin{figure}[t]  
\centering
\includegraphics[width=0.45\textwidth]{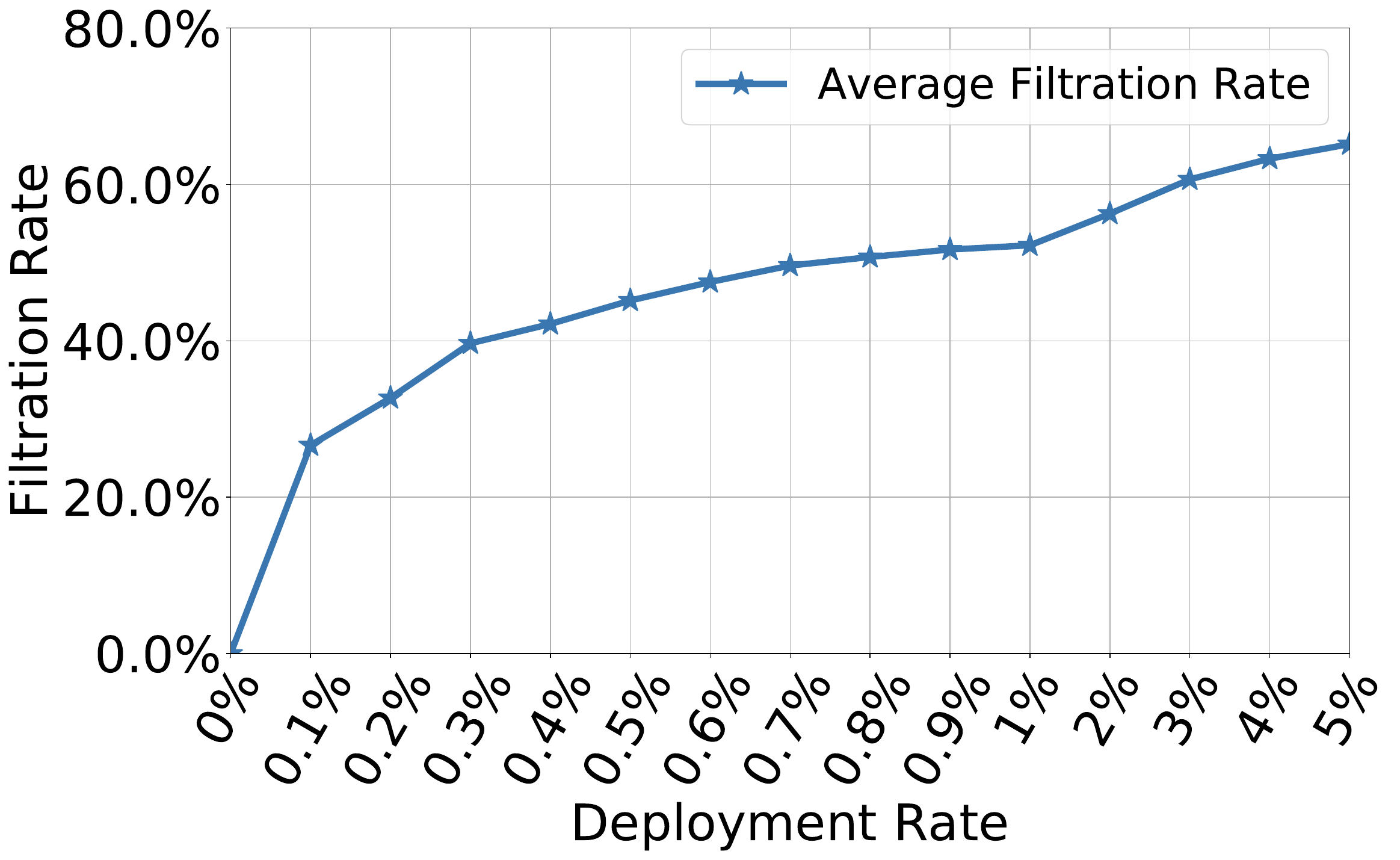}
\caption{Average Filtration Rate for Data Plane.}
\label{FiltrationRate}
\end{figure}

\parab{Quantitative Results} Using the same dataset as \S~\ref{sub:SecurityofCP} and the same set of deployment rates, we study the filtering rate $\mathcal{F}$ for all collected BGP paths. The average filtering rates are reported in Figure~\ref{FiltrationRate}. 
It is evident that as the deployment rate of \sys increases, the average filtering rate also increases. For instance, when the deployment rate reaches 0.7\%, 50\% of the unwanted traffic  can be discarded. We note that this analysis only considers the filtering by on-path ASes. In practice, the filtering rate is expected to be higher if we take into account collaborative filtering from off-path ASes. 


\subsection{Deployment}
\label{subsec:Deployment}

\parab{Slow Start in a Stable Network} 
In a well-established network, the number of new BGP updates is often much smaller than the total number of established BGP routes. For instance, the total number of BGP paths in the Caida dataset of April 2023 is about 377 million. The number of BGP paths remains the same over the period of 4 months (from January 2023 to April 2023) is about 260 million, roughly  68.99\% of all paths.  

Since new \fcs are generated only if BGP paths change, \sys may face a slow start issue that the  established BGP routes do not trigger \fcshort generation. Consequently, the subsequent data plane forwarding verification that relies on \fcshort binding relationships is also limited due to the lack of \fcs.  

As part of the initialization when an AS $N$ joins the \sys ecosystem, it shall generate an initial binding message for each of the local BGP routes. The version number $\textsf{Ver}$ in this message is set as a special indicator $0$, and the $\textsf{FC}_\textsf{list}$ is the inverse sequence of the AS-Path in the BGP route. The verification protocol of these messages is similar to regular the verification protocol (see \S~\ref{subsec:forward_binding}), except that no verification is needed for the  $\textsf{FC}_\textsf{list}$.

The startup binding messages are also helpful in detecting existing/established path manipulation or source spoof attacks. Suppose that from AS $S$'s perspective, its traffic $\langle \textsf{src}{:}S, \textsf{dst}{:}D\rangle$ expects to take the AS-path $S{\rightarrow}B{\rightarrow}C{\rightarrow}D$. Yet the actual forwarding path is manipulated as  $S{\rightarrow}M{\rightarrow}D$ to route through a compromised AS $M$. Once the traffic filters are constructed based on these startup messages, AS $D$ realize that traffic $\langle \textsf{src}{:}S, \textsf{dst}{:}D\rangle$ received from $M$ is unauthorized, and therefore should be discarded. 
Similarly, ASes can discard spoofed traffic based on these startup binding messages.

\parab{Initial Network} 
Contrary to the stable/established network is a newborn network without full connectivity yet. Thus, it is possible that some AS pairs do not have viable paths between them. 
To facilitate startup in a newborn network, we can rely on several well-connected service providers or CDNs to relay binding messages ASes. Specifically, instead of direct message broadcast, ASes can adopt a pub-sub mechanism to publish and pull binding messages from these service providers. 
Once the initialization phase is complete, \sys can switch back to the normal operation mode. 
\section{Evaluation}
\label{performance}

In this section, we evaluate the performance of FC-BGP.The implementation of FC-BGP and the design of the testbed are described in \S~\ref{subsec:testbed}.The experiments on the performance of the control plane are presented in \S~\ref{subsec:eval:control_plane}.The experiments on the performance of the propagation and consistency guarantees for data-plane binding messages are presented in \S~\ref{subsec:eval:dataplane}.


\subsection{Implementation and Testbed}
\label{subsec:testbed}

\begin{figure}[t]    
\centering
\includegraphics[width=0.4\textwidth]{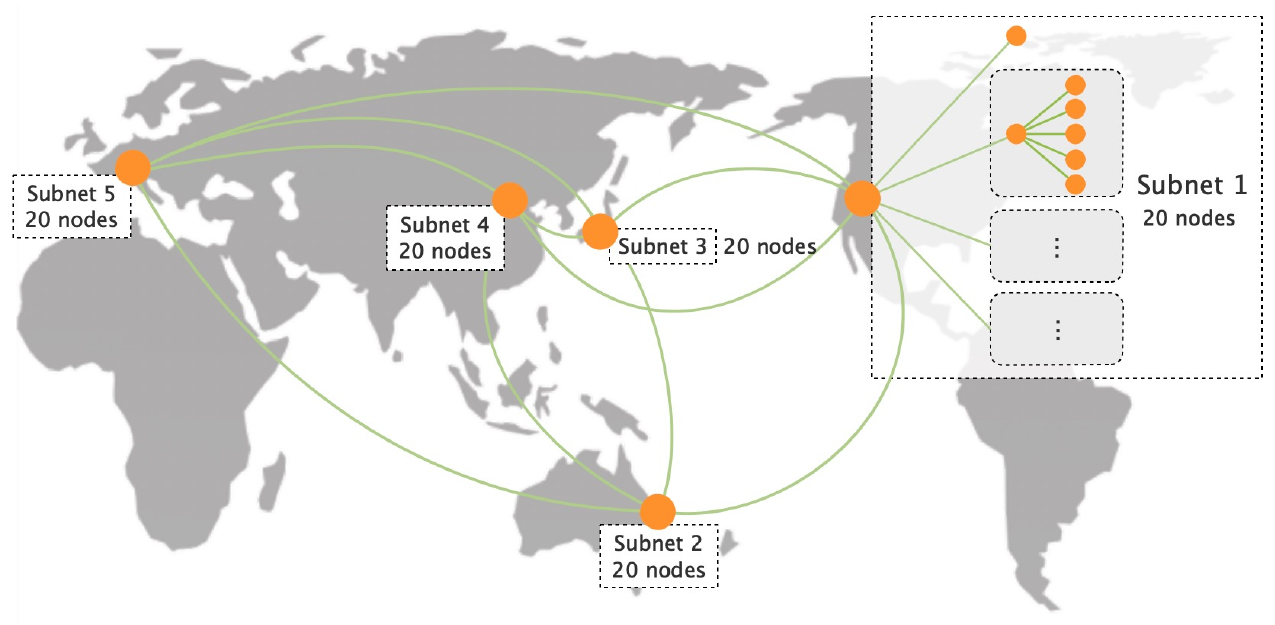}
\caption{The network topology for our global-scale testbed.}
\label{Topo}
\end{figure}

We implement control plane of \sys based on Quagga (version 1.2.4)\footnote{https://github.com/Quagga/quagga/releases/tag/quagga-1.2.4/}, and then further build a global-scale testbed over the global cloud infrastructure to evaluate \sys in the wild. The network topology is plotted in Figure~\ref{Topo}. We provision 100 virtual machines (each with a single core and 2G RAM) and evenly distribute them across five regions (Frankfurt, Beijing, Tokyo, Sydney, and Silicon Valley). Each region internally forms the same topology (a three-layer hierarchy), and is connected with all the other four external regions. The overall topology emulates the Internet AS hierarchy where there are several core ASes fully connected with other core ASes (forming peering relationships), and other lower-ranked ASes connect to their higher-ranked provider ASes (forming provider-customer relationships). 
We manually configure the BGP neighbor relationships (with the BGP multi-hop option enabled) on these nodes to form the desired overlay topology.

\subsection{Control Plane Overhead}
\label{subsec:eval:control_plane}

\begin{figure}[t]  
\centering
\includegraphics[width=0.48\textwidth]{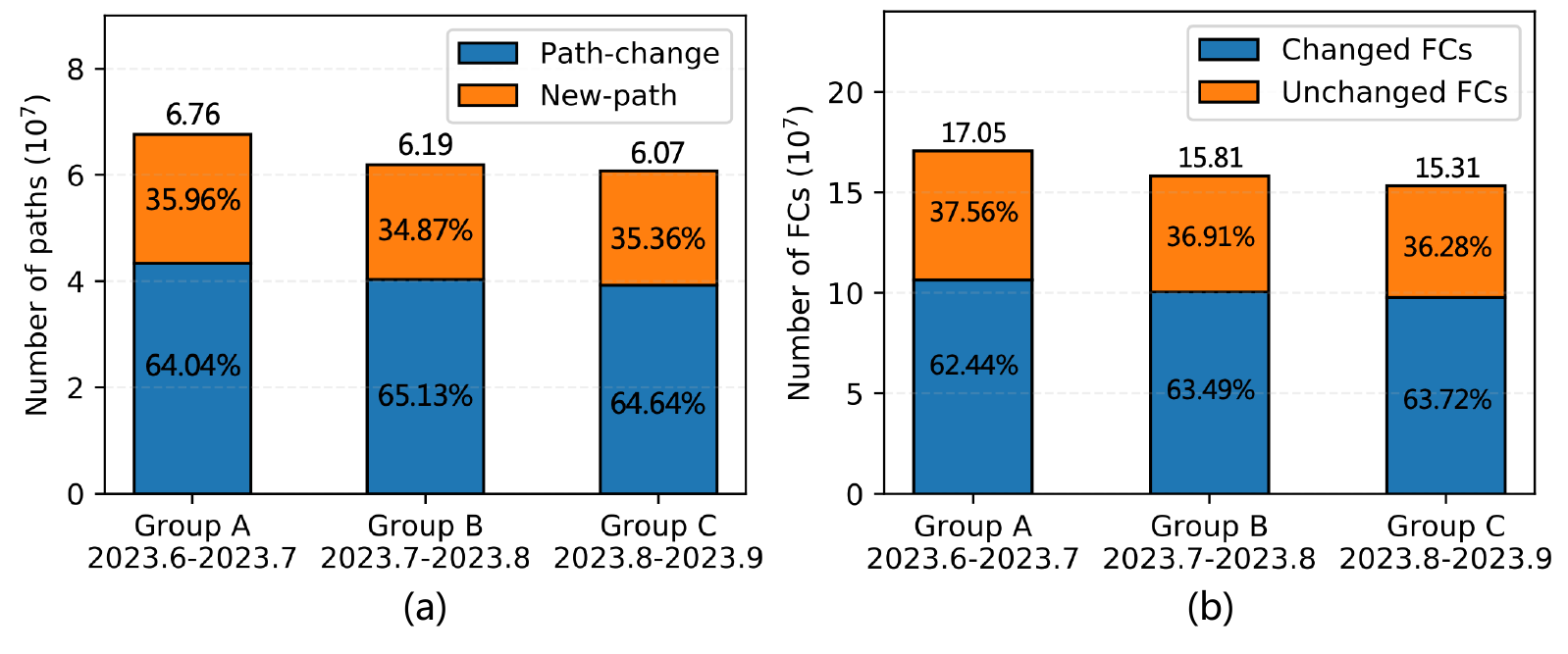}
\caption{Statistical results of the BGP updates.}
\label{paths-hops}
\end{figure}

\begin{figure}[t] 
\centering
\includegraphics[width=0.45\textwidth]{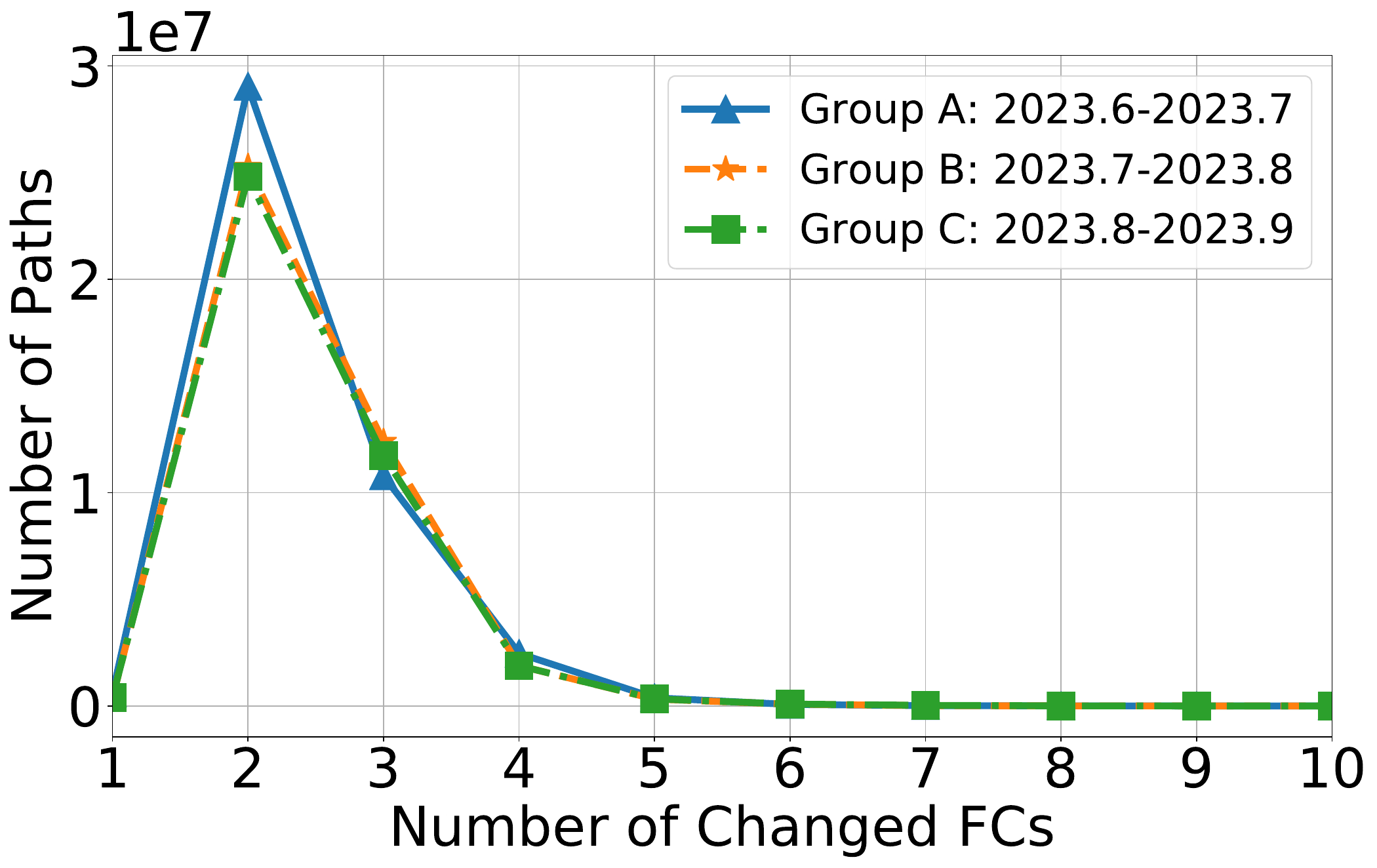}
\caption{The breakdown of path-change updates according to the number of changes \fcs.}
\label{paths-dis}
\end{figure}

\subsubsection{Internet-Scale BGP Update Validation}
\label{subsubsection:Internet Validation}
Because \fcs are hop-specific, whenever a BGP path changes, \sys only needs to verify the \fcs generated for those updated hops. On the contrary, since the verification of BGPsec is bound to a specific path, it needs to verify the entire new path. 
To quantify the overhead difference for these two different verification schemes, we analyze detailed BGP route updates in the CAIDA dataset. 
Specifically, based on the BGP announcement dataset collected from January to April 2023, we analyze the BGP path changes over three one-month period, \ie Group A represents the path changes happened from January to February, Group B represents the changes from February to March, and Group C represents the changes from March to April. 
We define the following metrics to represent BGP path changes. 
\begin{itemize}[leftmargin=*]
    \item \emph{New-path update} represents a BGP update with a new (src-AS, prefix) tuple that is not received before. 

    \item \emph{Path-change update} represents a BGP update with an existing (src-AS, prefix) tuple, yet its AS-path differs from that of a previously received BGP update with the same (src-AS, prefix) tuple. 
    We also record the number of changed \fcs between the two AS-paths. 
\end{itemize}

We plot the results in Figure~\ref{paths-hops}. The experimental results show that in each group, over 60\% of BGP updates are path-change updates (Figure~\ref{paths-hops}(a)), within which over 36\% of \fcs remain the same (Figure~\ref{paths-hops}(b)).This means that at least 21.6\% of \fcs do not need to be re-validated. For instance, among the 67 million BGP updates in Group A, over 43 million updates are path-change updates, within which 37.56\% of the pathlets remain unchanged. As a result, \sys can save a non-trivial amount of cryptography signing and verification compared with BGPsec. 



In Figure~\ref{paths-dis}, we further 
categorize all the path-change updates according to the number of changed \fcs. In most cases, a one-hop change in path will cause two \fcs to change. There are only a few cases where a one-hop change causes one \fcshort change, such as when $\mathcal{P}{:} A {\leftarrow} B {\leftarrow} C$ turns into $\mathcal{P}{:} A {\leftarrow} B {\leftarrow} D$.
The experimental results shows that the path-change updates are dominated by the case where only a small number of \fcs changed. Specifically, over 64\% path-change updates include no more than 2 \fcs changes. This implies that the pathlet-based verification in \sys is highly efficient in practice. 

Meanwhile, we deploy \sys on our large-scale overlay network and measure the actual control plane verification latency. The result shows that the extra delay introduced by \sys for BGP verification is significantly smaller than transmission delays. Limited by space, please see \S~\ref{100latency} for more detail.

\subsection{Dataplane Validation Overhead}
\label{subsec:eval:dataplane}

\subsubsection{Binding Message Synchronization}
\label{subsubsec:eval:sync_eff}
As discussed in \S~\ref{subsubsec:version_view}, instead of employing existing BFT protocols to agree upon the dataplane binding messages globally, \sys designs a fast binding message publication protocol by decoupling message broadcast and periodic consistency check. 
In this segment, we quantitatively compare the efficiency of \sys with the partially synchronous PBFT protocol~\cite{castro1999practical}. PBFT  is often adopted as the building block to optimize the consensus in large-scale blockchain platforms like FISCO BCOS~\footnote{https://www.fisco.org.cn}. 

\begin{figure}[t]    
\centering
\includegraphics[width=0.45\textwidth]{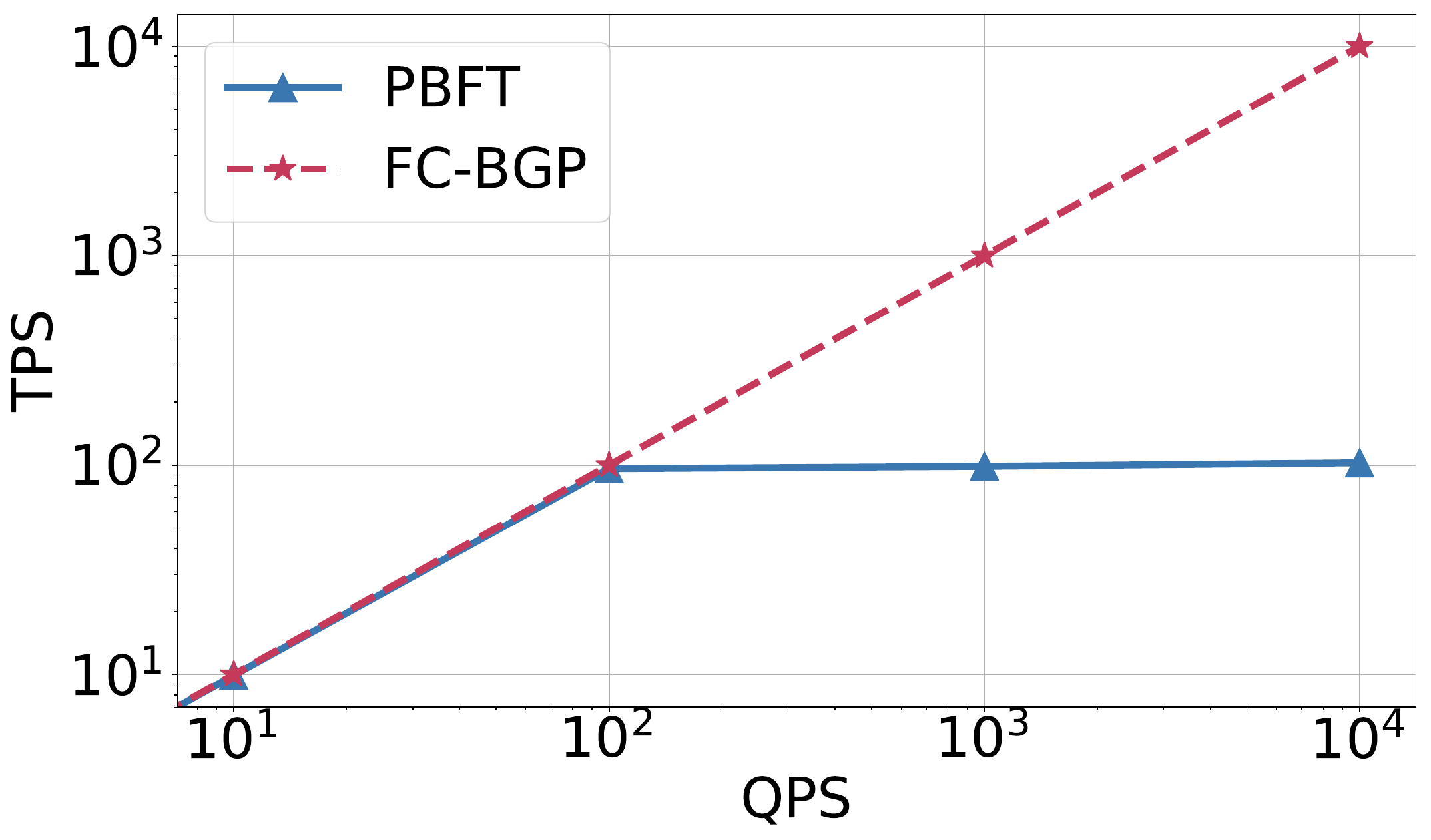}
\caption{Throughput comparison between our consistency check and PBFT.}
\label{Throughput}
\end{figure}

We first evaluate the throughput of PBFT and our protocol using the 100-AS global network. 
We measure the maximum transaction per second (TPS) achievable by both protocols as we increase the number of binding messages (denoted as queries per second, QPS) generated by each AS. The results, plotted in Figure~\ref{Throughput}, show than the TPS of PBFT quickly saturated as the QPS reaches 100, while the TPS of our protocol scales up to handle 10000 QPS per AS (one million QPS in aggregate). This is because PBFT needs to broadcast binding messages to all nodes and then needs to wait for the leader to pack all these messages before initiating the consensus. 
This process consumes significant bandwidth, which can quickly overflow the capacity of a wide area network. In contrast, our protocol only broadcasts lightweight version vectors to each node for consistency check. This significantly reduces the bandwidth consumption, allowing our protocol to scale to much higher TPS levels.

\begin{figure}[t]    
\centering
\includegraphics[width=0.45\textwidth]{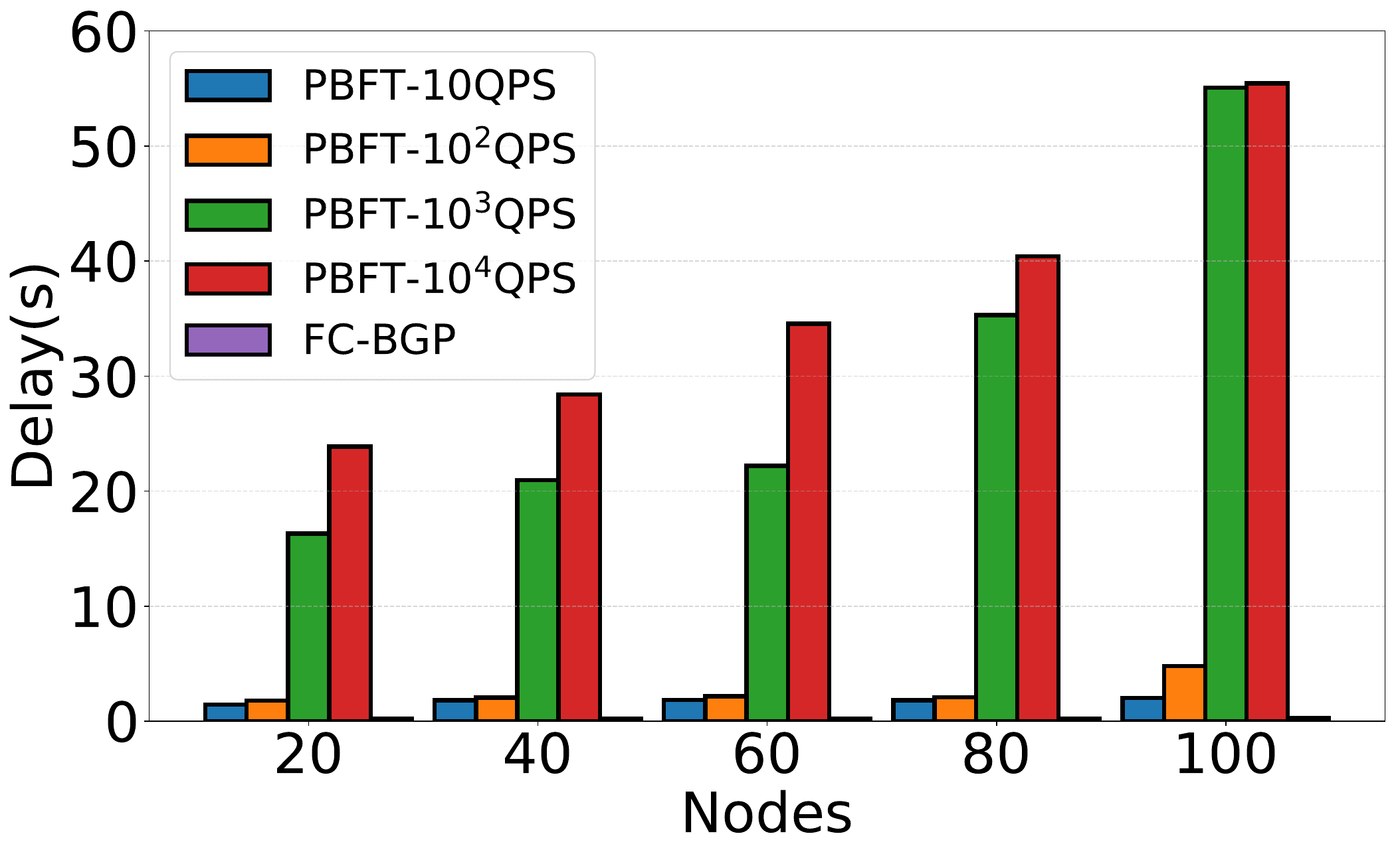}
\caption{Latency comparison between our consistency check and PBFT.}
\label{Con-test-RP}
\end{figure}

In addition to the throughput, we further evaluate the synchronization/consensus delay of \sys and PBFT. 
The experiments are divided into five parts, each with a different number of nodes: 20, 40, 60, 80, and 100. The nodes in each part were evenly distributed across five regions. In each part of the experiment, we measured the delay of PBFT under different QPS rates: 10, 100, 1000, and 10000. The results are plotted in Figure~\ref{Con-test-RP}. We fix QPS as 10000 per AS for \sys. 
We observed that even when PBFT is not overloaded by high QPS rates, its consensus delay is about 7 to 20 times that of \sys. This indicates that even if it were possible to provide the network with higher bandwidth, the consensus delay of PBFT would still be significant. This essentially makes it difficult to maintain the stability of the system, as we discuss below.
When PBFT is overloaded, its delay grows significantly as expected.

\subsubsection{System Stability}\label{subsubsec:steady}

\begin{figure}[t]    
\centering
\includegraphics[width=0.45\textwidth]{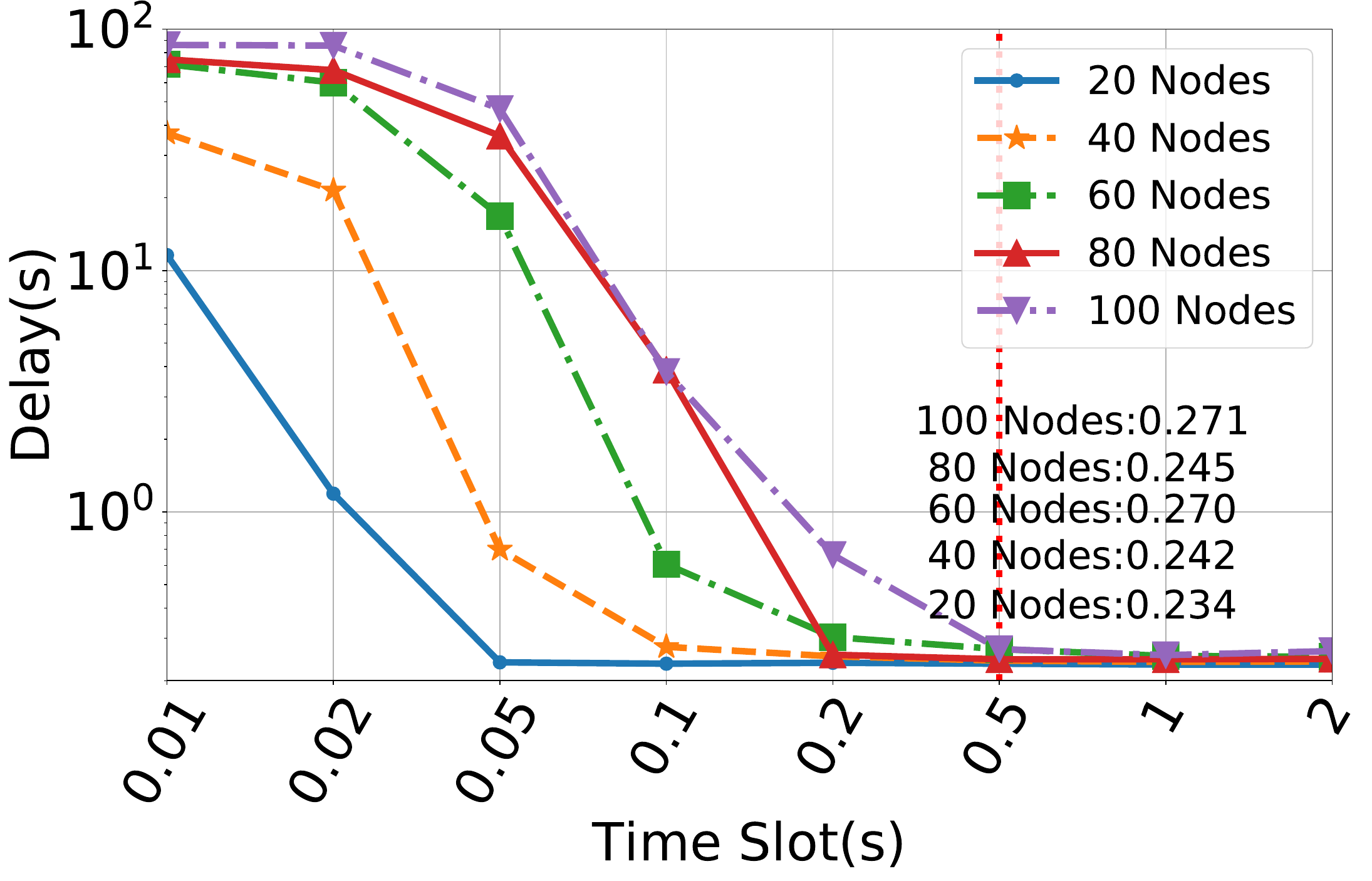}
\caption{System stability}
\label{Con-test}
\end{figure}

To ensure stability, the average consistency check latency should be less than the consistency check period (otherwise the binding messages would accumulated and eventually overload the system). In this section, we explore that the stable rate at which \sys should periodically check the binding messages. We consider eight different check periods, ranging from 0.01 to 2 seconds, and 5 different network scales. For each period-scale combination, we repeat the experiment 1000 times to avoid statistical biases. The experimental results are plotted in Figure~\ref{Con-test}. 

The results show that when the consistency check period is short, the consistency check latency explodes because many binding messages are queued up. The smaller the number of ASes, the shorter the consistency check period at which our synchronization protocol can run stably. 
For the 100-AS network, \sys is stable as long as the consistency check period is greater than 0.5 second, while for the 80-AS network, \sys is stable as long as the period is greater than 0.2 second. 
The inter-region network transmission delay (which is on the orders of hundreds of milliseconds) is the theoretical lower bound at which the consistency check can execute. This is because it takes time for the binding messages to be transmitted between regions. In our system, the actual bound is quite close to the theoretical bound.


\section{Related Work}
\noindent \textbf{Security Enhanced BGP.} 
BGP security has been extensively studied~\cite{butler2009survey} and in this section, we will focus on how to prevent path manipulation attacks. As the first BGP security solution, S-BGP introduced encryption mechanisms to BGP validation for the first time. Several variations~\cite{hu2004spv,zhao2005aggregated,butler2006optimizing,xiang2013sign,aiello2003origin,cohen2015one,raghavan2007analysis, soBGP,psBGP} has been developed to reduce the overhead of S-BGP. For example, soBG~\cite{soBGP} utilizes EntityCert and web of trust to accomplish BGP path validation, while psBGP~\cite{psBGP} abandons S-BGP's management of address certificates and introduces evaluation of AS to achieve distributed path authentication. In addition, from a more practical perspective, Path-End Validation~\cite{cohen2016jumpstarting} proposes an extension of RPKI to provide effective BGP path protection through a small number of validations. In another category of research, the exploration to enhance BGP security based on block-chain has also yielded positive results~\cite{he2020roachain,paillisse2018ipchain,de2017blockchain,xing2018bgpcoin}. Such research is currently focused on the area of origin verification, and additionally faces challenges in terms of verification latency, privacy protection, and energy waste.

Meanwhile, the IETF SIDR working group proposed the Reserve Public Key Infrastructure (RPKI)\cite{RPKI}, Route Origin Authorization(ROA), and BGPsec\cite{BGPsec} on the basis of S-BGP. As the most influential solutions in current Internet practice, RPKI, ROA and BGPsec carried out a lot of work around the above mechanisms, such as security enhancements\cite{cooper2013risk,heilman2014consent} and deployment measurements\cite{iamartino2015measuring,wahlisch2012towards}, etc.





\noindent \textbf{Packet Forwarding Verification.} 
A secure forwarding process implies forwarding packets from the real source to the destination following a reliable path announced by the BGP and, at the same time, protecting the packets from tampering in the process. Packet tamper-proofing can be achieved by various cryptography-based methods \cite{kent2005security,kent2005rfc,bremler2005spoofing,liu2008passport,kim2014lightweight,legner2020epic,wu2018enabling}. 


There has been a category of source address validation schemes that utilize routing information to determine legitimate packet flow and thus establish filtering rules. For example, Distributed Packet Filtering(DPF) \cite{park2001effectiveness}, Inter-domain Distributed Packet Filter(IDPF) \cite{duan2008controlling}, ingress/egress filtering also known as BCP 38 \cite{ferguson2000defeating}, Unicast Reverse Path Forwarding(uRPF) \cite{kumari2009remote}, and enhanced feasible-path uRPF(EFP-uRPF) \cite{baker2004rfc3704, sriram2020enhanced}. However, this class of mechanisms all relies in different ways on the security and reliability of the routing information itself, and this insufficiently robust foundation largely affects the effectiveness of the mechanisms involved.

The authenticity assurance of forwarding paths has also been extensively researched for more than a decade. A prominent contribution of the early research was to provide an important design paradigm~\cite{savage2000practical,yaar2003pi,yaar2006stackpi}. With a key negotiated between the host and the routing device, the device embeds verifiable information into the packet header during forwarding, which is used at the receiving end to recover the actual path and thus determine the authenticity of the forwarded path. Several subsequent researches, such as ICING~\cite{naous2011verifying}, OPT~\cite{kim2014lightweight}, EPIC~\cite{legner2020epic}, OSP~\cite{cai2015source} and PPV~\cite{wu2018enabling}, are devoted to optimizing the serious overhead problem on this technical route while improving the security of the scheme by various ways (e.g., symmetric encryption, probabilistic verification, homomorphic encryption).

\section{Conclusion}

In this paper, we propose \sys, a novel secure inter-domain routing system that can simultaneously authenticate BGP announcements and validate dataplane forwarding in an efficient and incrementally-deployable manner. The core  design of \sys is a publicly verifiable code named Forwarding Commitment. \fcshort certifies an AS's routing intent on one of its directly connected neighbors. Centering around the \fcs, we design \first a BGP announcement authentication mechanism that achieves the same security guarantees as BGPsec while offering significantly more security benefits in case of partial deployment; and \second a dataplane forwarding validation mechanism that can enable ASes, both on-path and off-path, to discard unwanted traffic that deviates from its authorized forwarding path on the data plane. We implement a prototype of \sys and extensively evaluate it both analytically and experimentally.

\bibliographystyle{plain}
\bibliography{reference}

\balance
\appendix
\section{Appendix}
\label{appendix}

\subsection{Propagation of \fcs}
\label{p-fc}

\begin{figure}[t]    
\centering
\includegraphics[width=0.395\textwidth]{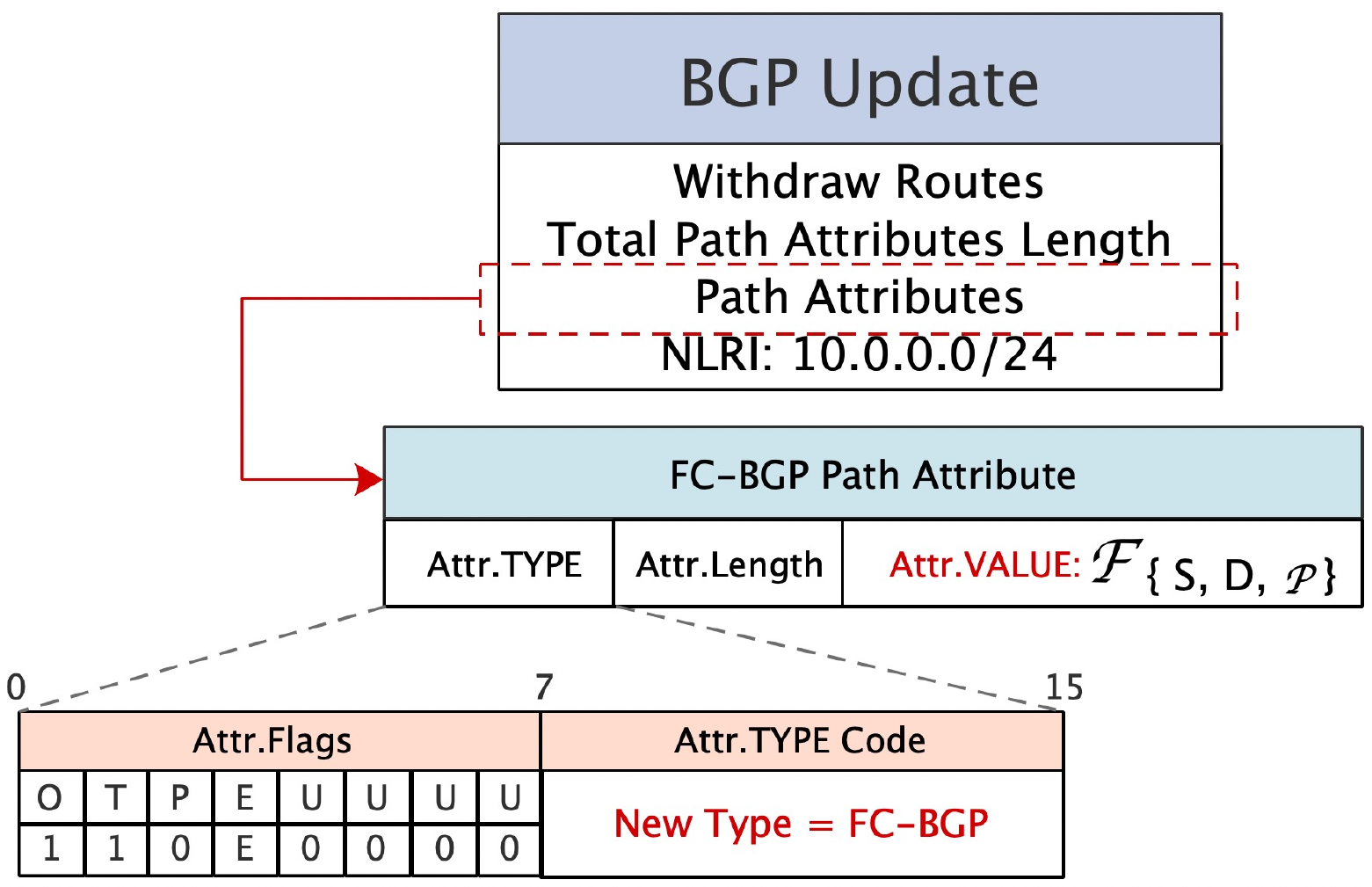}
\caption{BGP Update Message with FC}
\label{FCPropagation}
\end{figure}

\fcs are propagated along with BGP update messages via a newly allocated path attribute type. In Figure~\ref{FCPropagation}, we plot the format of the BGP update message carrying the \fcs. The \fcshort path attribute is in Type-Length-Value format, where \textsf{Attr.TYPE} identifies the class of the current path attribute, which consists of Flags and Code. 
\textsf{Attr.Flags} in total occupies one byte: the O (Optional) bit and T (Transitive) bit, indicating the \fcshort path attribute is optional and transitive. According to \cite{rekhter2006border}, we set the P (Partial) bit as 1 to ensure that the \fcshort attribute can be propagated even by legacy ASes (\ie they do not recognize the \fcshort attribute). 
The E (Extended Length) bit is set on demand according to the length of AS-path: it is set to 1 if \textsf{Attr.Length} is greater than 1 (\ie this BGP update carries more than one \fcshort). The remaining four bits in \textsf{Attr.Flags} are unused. 
\textsf{Attr.TYPE Code} represents the new path attribute code, which should not be collide with existing codes published by IANA \footnote{https://www.iana.org/assignments/bgp-parameters/bgp-parameters.xhtml}. In our prototype, we set this to $41$ which is not yet assigned. Finally, the \textsf{Attr.VALUE} contains all \fcs that have been added in this message. 

\subsection{Topology on the Overlay Network}
\label{100latency}
\begin{figure}[t]    
\centering
\includegraphics[width=0.35\textwidth]{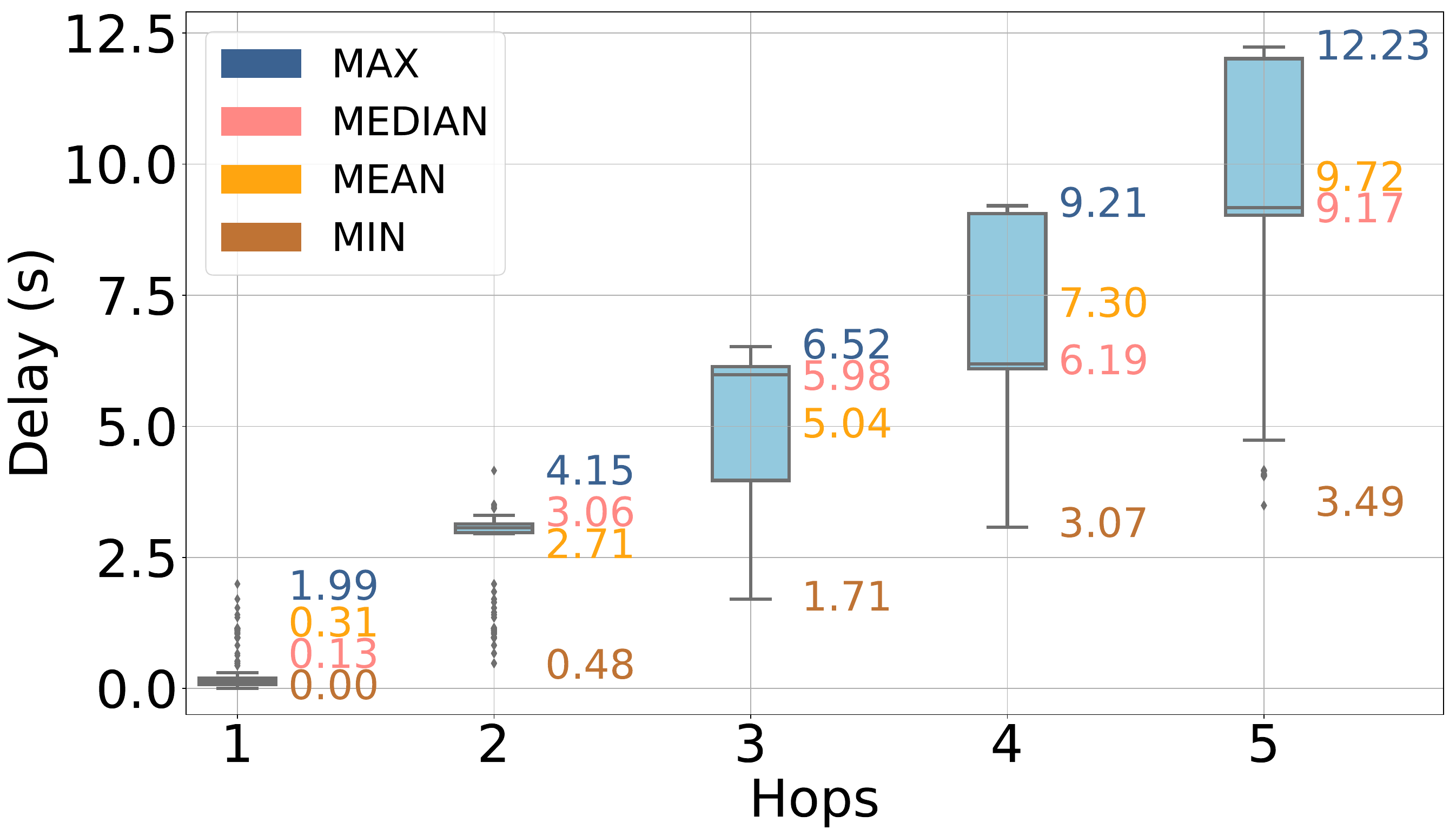}
\caption{The control plane latency of \sys on our large-scale overlay network.}
\label{Delay-100}
\end{figure}

We deploy \sys on our large-scale overlay network and measure the actual control plane verification latency. 
We enable \sys on all 100 nodes and let them start sending and receiving BGP announcements to all neighbors simultaneously, simulating a large-scale network with 100 ASes distributed globally. 
For each announcement, a node records the following tuple: $ \{action | hops | pre\_AS | src\_AS |ts \}$, where $action$ indicates sending or receiving, $hops$ records the current hop count in the announcement, $pre\_AS$ and $src\_AS$ record the previous AS and source AS, respectively, and $ts$ records the time of this event. 
When the network converges (\ie no new BGP updates are generated for 30 minutes), we aggregate the measurement tuples records on all 100 ASes to obtain the latency statistics for processing each BGP announcement, including the maximum, minimum, median and average processing times. 
We plot the results in Figure~\ref{Delay-100}. 

\begin{table}[t]
  \begin{center}
    \caption{Delay Comparison}
    \resizebox{\linewidth}{!}{
    \begin{tabular}{c|c|c|c} 
      \textbf{} & \textbf{Average} & \textbf{Max}& \textbf{Min}\\
      \hline
      Intra-region delay & 14.85ms & 35.3ms & 35.3ms\\
      Inter-region delay & 130.12ms & 599.02ms & 42.79ms\\
      \sys processing delay & 0.025ms & 0.23ms & 0.013ms\\
    \end{tabular}}
    \label{delay-com}
  \end{center}
\end{table}
\label{delay}

To put the control plane overhead of \sys in perspective, we list the network transmission delays and the times that it takes \sys to generate and verify forwarding commitments in Table~\ref{delay-com}. Clearly, the extra  delay introduced by \sys for BGP route verification is significantly smaller than transmission delays (roughly 600x and 5200x smaller than the intra-region and inter-region transmission delays, respectively). 
Therefore, in large-scale deployments, the impact of small jitters in transmission delays will far outweigh the overhead incurred by \sys.


\end{document}